\documentclass{aastex62}
\usepackage{amssymb}
\usepackage{graphicx,graphics}													
\usepackage{epsfig}																
\usepackage{dcolumn}																
\usepackage{bm}																	
\usepackage{natbib}
\usepackage{ulem}
\usepackage{times}

\newcommand{\beq}{\begin{equation}}
\newcommand{\eeq}{\end{equation}}

\newcommand{\hi}{H{\sc i}}
\newcommand{\hii}{H{\sc i}~21\,cm}

\shorttitle{Probing star formation in galaxies at $z \approx 1$}
\shortauthors{Bera et al.}
\begin{document}
\title{Probing star formation in galaxies at  $z \approx 1$ via a Giant Metrewave Radio Telescope stacking analysis}

\correspondingauthor{Nissim Kanekar}
\email{nkanekar@ncra.tifr.res.in}

\author{Apurba Bera}
\affiliation{National Centre for Radio Astrophysics, Tata Institute of Fundamental Research, Pune University, Pune - 411007, India}

\author{Nissim Kanekar}
\altaffiliation{DST Swarnajayanti Fellow}
\affiliation{National Centre for Radio Astrophysics, Tata Institute of Fundamental Research, Pune University, Pune - 411007, India}

\author{Benjamin J. Weiner}
\affiliation{Steward Observatory, Department of Astronomy, University of Arizona, Tucson, AZ 85721}

\author{Shiv Sethi}
\affiliation{Department of Astronomy and Astrophysics, Raman Research Institute, C.V Raman Avenue, Bengaluru, India}

\author{K. S. Dwarakanath}
\affiliation{Department of Astronomy and Astrophysics, Raman Research Institute, C.V Raman Avenue, Bengaluru, India}

\begin{abstract}
We report deep Giant Metrewave Radio Telescope (GMRT) 610 MHz continuum imaging of four sub-fields of 
the DEEP2 Galaxy Redshift Survey. We stacked the radio emission in the GMRT
images from a near-complete (absolute blue magnitude ${\rm M_B} \leq -21$) sample of 3698 blue 
star-forming galaxies with redshifts $0.7 \lesssim z \lesssim 1.45$ to detect (at $\approx 17\sigma$ 
significance) the median rest-frame 1.4~GHz radio continuum emission of the sample galaxies. The 
stacked emission is unresolved, with a rest-frame 1.4~GHz luminosity of 
$\rm L_{1.4 \; GHz} = (4.13 \pm 0.24) \times 10^{22}$~W~Hz$^{-1}$. We used the local relation between 
total star formation rate (SFR) and 1.4~GHz luminosity to infer a median total SFR of 
$\rm (24.4 \pm 1.4)\; M_\odot$~yr$^{-1}$ for blue star-forming galaxies with 
$\rm M_B \leq -21$ at $0.7 \lesssim z \lesssim 1.45$. 
We detect the main-sequence relation between SFR and stellar mass, $\rm M_\star$, obtaining 
$\rm SFR = (13.4 \pm 1.8) \times [(M_{\star}/(10^{10} \;M_\odot)]^{0.73 \pm 0.09} \; M_\odot \; yr^{-1}$; 
the power-law index shows no change over $z \approx 0.7 - 1.45$. We find that the nebular line emission 
suffers less extinction than the stellar continuum, contrary to the situation in the local Universe;
the ratio of nebular extinction to stellar extinction increases with decreasing redshift. We obtain an 
upper limit of $0.87$~Gyr to the atomic gas depletion time of a sub-sample of DEEP2 galaxies at 
$z \approx 1.3$; neutral atomic gas thus appears to be a transient phase in high-$z$ star-forming galaxies.
\end{abstract}

\keywords{galaxies: high-redshift --- galaxies: star formation --- radio continuum: galaxies}

\section{Introduction} 
\label{sec:intro}

In recent years, optical imaging and spectroscopic studies of the ``deep fields'' 
\citep[e.g. the Hubble Deep Fields, the Chandra Deep Field South, the COSMOS field, 
etc; e.g. ][]{dickinson03,giavalisco04,scoville07} have yielded detailed information on the star 
formation activity in galaxies over a wide range of redshifts \citep[e.g.][]{madau14}. 
Such studies have shown that the comoving star formation rate (SFR) density rises 
steadily from $z \approx 7$ to $z \approx 3$, is roughly flat over $z \approx 1-3$, and then 
declines by a factor of ten from $z \approx 1$ to the present epoch 
\citep[e.g.][]{lefloch05,hopkins06,bouwens14}. Both the metallicity and the SFR of star-forming
galaxies have been found to depend on stellar mass, in the local Universe and at high 
redshifts \citep[e.g.][]{tremonti04,brinchmann04,erb06,noeske07}. 

The tight relation between SFR and stellar mass in star-forming galaxies (the ``main sequence'', 
with SFR~$\rm \propto M_\star^\alpha$) and its redshift evolution have been topics of extensive 
scrutiny over the last decade, with a variety of studies using galaxy samples selected based on 
different criteria and using different SFR indicators 
\citep[e.g.][]{brinchmann04,salim07,daddi07,noeske07,elbaz07,pannella09,pannella15,santini09,karim11,whitaker14,tasca15}.
The power-law index and normalization of the main sequence have been shown to play important roles 
in the evolution of galaxies and their mass functions \citep[e.g.][]{renzini09,peng10}.

However, most studies of the main sequence in high-$z$ galaxies are based on SFR estimates from 
optical imaging or spectroscopy, using rest-frame ultraviolet (UV) or optical tracers. These are 
susceptible to dust obscuration effects, and typically only yield the unobscured SFR, uncorrected
for dust extinction. Conversely, radio continuum studies allow one to determine the total SFR of galaxies, 
from both unobscured and obscured star formation, from the tight correlation between the 1.4~GHz radio 
luminosity and the far-infrared (FIR) luminosity of star-forming galaxies 
\citep[e.g.][]{condon92,yun01,pannella15,magnelli15}. Calibration relations can then be used to 
infer the total SFR of a galaxy from its measured rest-frame 1.4~GHz radio luminosity 
\citep[e.g.][]{yun01,bell03,kennicutt12}. 

Unfortunately, deep integrations with current radio telescopes are needed to detect the rest-frame 
1.4~GHz continuum emission from normal star-forming galaxies at even low redshifts, $z \approx 0.2$. 
However, it is possible to infer the {\it statistical} star formation properties of a sample of 
galaxies by stacking their 1.4~GHz continuum emission, and thence deriving their median (or mean)
SFR from the stacked 1.4~GHz radio luminosity \citep[e.g.][]{white07}. Such studies can then be 
used to trace the dependence of the median SFR on various galaxy properties, such as metallicity, 
stellar mass, color, redshift, etc.


\begin{table*}[t]
\centering
\caption{Summary of the observational details and results}
\begin{tabular}{p{3cm} c c c c }
\hline
			&	Sub-field 1	    &	Sub-field 2		  &	Sub-field 3	      &	Sub-field 4 \\
\hline \\
RA (J2000)		&  16h~48m~00.0s	    &  16h~51m~00.0s	 	  &  23h 28m 00.0s	      &	23h 32m 00.0s	        \\
DEC (J2000)		&  $34^{\circ}~56'~00.0''$  &  $34^{\circ} ~56' ~00.0''$  & $0^{\circ} ~ 9' ~ 00.0''$ & $0^{\circ}~ 9'~00.0''$  \\
Central frequency (MHz)	&	617.73		    &   617.73		          &	637.73		      &	637.73		        \\
On-source time (Hrs)    &        8.5                &    8.5			  & 	13                    & 13			\\
Synthesized beam	&  $4.7'' \times 3.9''$	    &	$5.2''\times 4.3''$	  & $5.9'' \times 4.6''$      &	$6.1'' \times 4.4''$	\\
RMS noise ($\mu$Jy)	&	21		    &	39			  &	22		      &	14			\\
Stacked galaxies	&	987		    &	994			  &	936		      &	781			\\	
\\ \hline
\end{tabular}
\label{table:fields}
\end{table*}

Most radio stacking studies in the literature use 1.4~GHz radio continuum images of optical deep fields
\citep[e.g.][]{carilli08,dunne09,pannella09,pannella15}. While this implies a high continuum sensitivity, 
the stacking for high-$z$ galaxies is effectively being done at a far higher {\it rest-frame} frequency than 
1.4~GHz. One has to then assume a spectral index for the radio emission to infer the rest-frame 1.4~GHz 
luminosity, and thence, the total SFR. We report here deep Giant Metrewave Radio Telescope (GMRT) 610~MHz 
continuum imaging of four sub-fields of the DEEP2 Galaxy Redshift Survey \citep{newman13}, which allow 
us to probe the dependence of the SFR on redshift and stellar mass for a near-complete sample of 
star-forming galaxies at $0.7 \lesssim z \lesssim 1.45$ via stacking of their rest-frame $1-1.4$~GHz 
radio continuum emission.\footnote{We will assume a 
flat $\rm \Lambda$-cold-dark-matter cosmology, with $\rm H_0=67.8\;km\;s^{-1}\;Mpc^{-1}$, 
$\rm \Omega_m=0.31$ and $\rm \Omega_\Lambda = 0.69$ \citep{planck16}.}

\section{The DEEP2 survey fields: GMRT observations and data analysis}
\label{sec:deep2}

The DEEP2 fields \citep{newman13} were chosen as the targets for our GMRT 610~MHz observations 
\citep[][]{kanekar16}. The DEEP2 Survey provides spectroscopic redshifts, using the O{\sc ii}$\lambda$3727 
doublet, for $\approx $ 38000 galaxies at $z \approx 0.7-1.45$ over 2.8~$\rm deg^2$ area on the sky, and 
is complete to an apparent magnitude of $\rm R_{AB}=24.1$ \citep[i.e. an absolute B-band magnitude of 
$\rm M_B=-20$ at $z\approx 1$; ][]{newman13}. The values of $\rm M_B$ and $\rm U-B$ from the DEEP2 Survey 
were used to estimate the stellar masses of the galaxies, assuming a Salpeter initial mass function 
with the calibration of \citet{weiner09}. Only galaxies with spectroscopic redshifts 
of quality 3 or 4 \citep[i.e. ``secure'' redshifts, with $\geq 95$\% 
probability of being correct; see ][]{newman13} were included in our analysis.

The GMRT 610~MHz receivers were used in 2012 to observe four of the DEEP2 sub-fields, with a bandwidth 
of 33.33~MHz sub-divided into 512~channels. Two sub-fields were observed at a central frequency of 
617.73~MHz and two at 637.73~MHz, with total times of $\approx 12-18$~hours per sub-field \citep{kanekar16}. 
The GMRT primary beam has a full width at half maximum (FWHM) of $\sim 43'$ at these frequencies, 
which covers an entire DEEP2 sub-field ($36'\times30'$) in a single pointing. 

The initial data analysis, including data editing, initial calibration, and self-calibration, was carried 
out in ``classic'' AIPS \citep[and is described in detail in][]{kanekar16}. The self-calibrated visibilities were then imaged in 
CASA, using the w-projection algorithm \citep{cornwell08}, to produce the final continuum images (including 
a correction for the shape of the GMRT primary beam). Table~\ref{table:fields} summarizes the observational 
details and results for the four GMRT pointings. These are amongst the deepest radio continuum images at 
$\approx 610$~MHz in the literature \citep[e.g.][]{taylor16}.

\begin{figure}[b]
\centering
\includegraphics[scale=0.79,trim={0.6cm 2cm 0.5cm 1.0cm},clip]{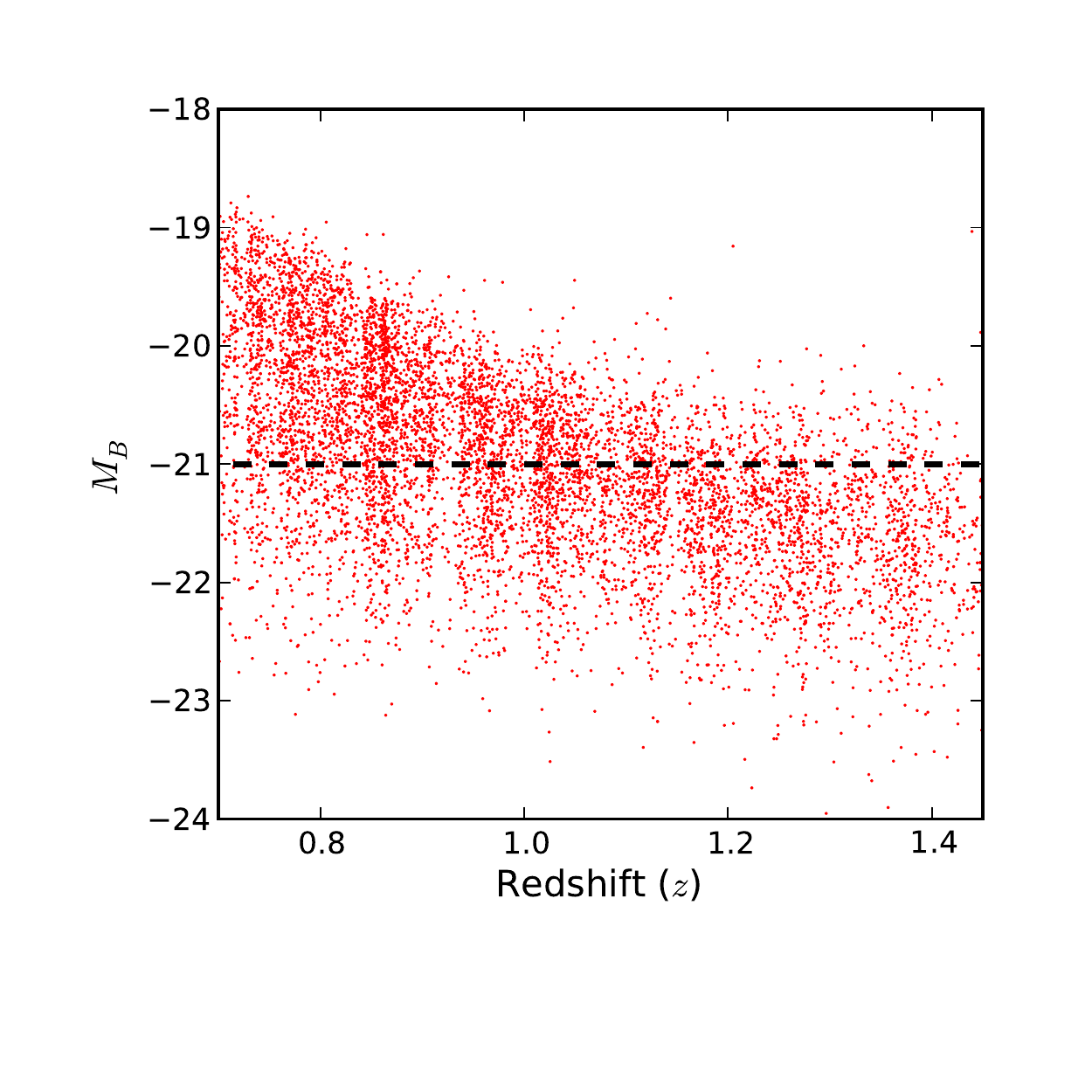}
\caption{Rest frame absolute B band magnitudes of the DEEP2 galaxies within the FWHM of the GMRT images. 
We retain galaxies with $\rm M_B\leq -21$ in our stacking analysis, to obtain a complete sample 
out to $\rm M_B \leq -21$ over our entire redshift range. 
\label{fig:mb_z}}
\end{figure}

\begin{figure*}[t]
\centering
\includegraphics[scale=0.35,clip]{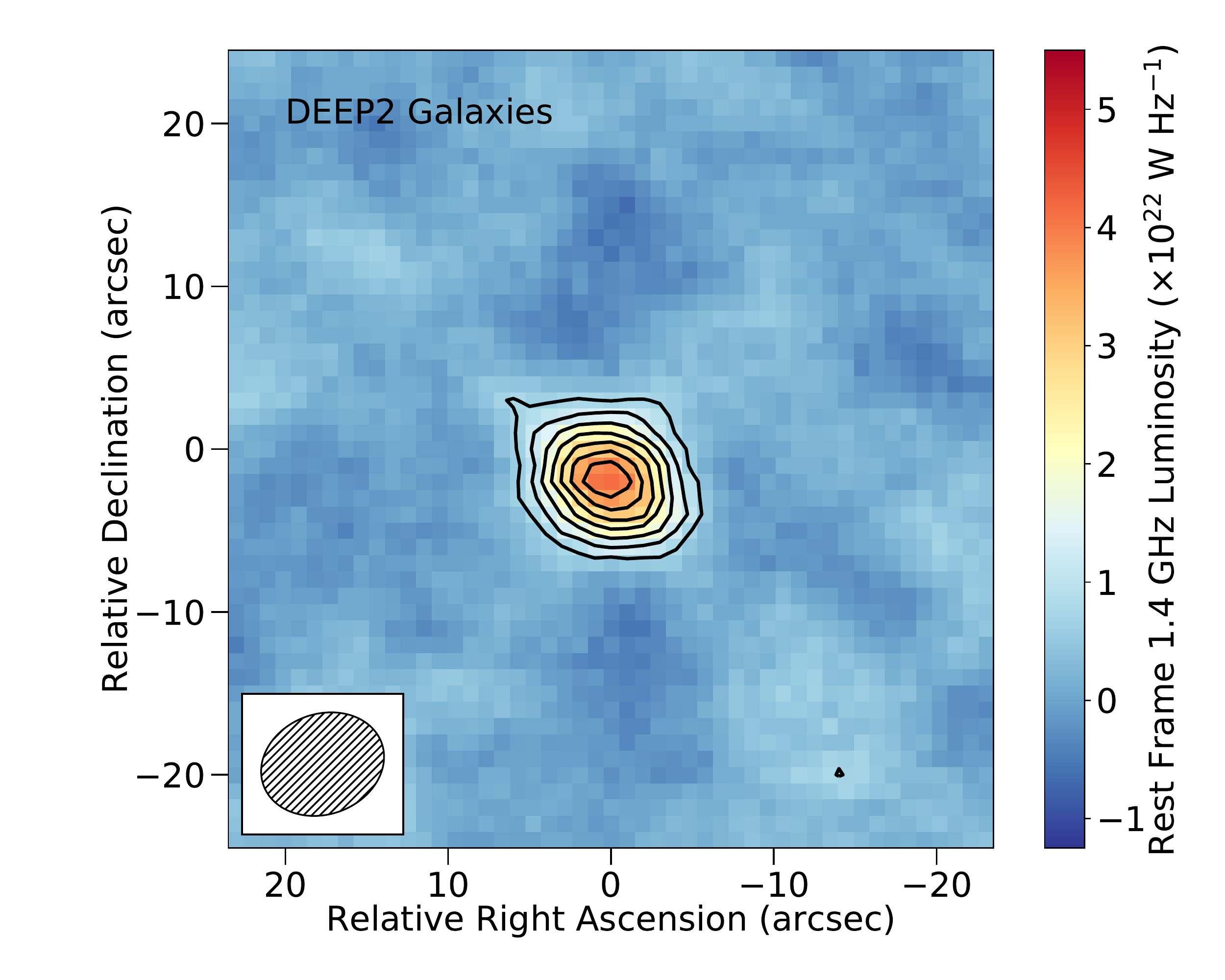}
\includegraphics[scale=0.35,clip]{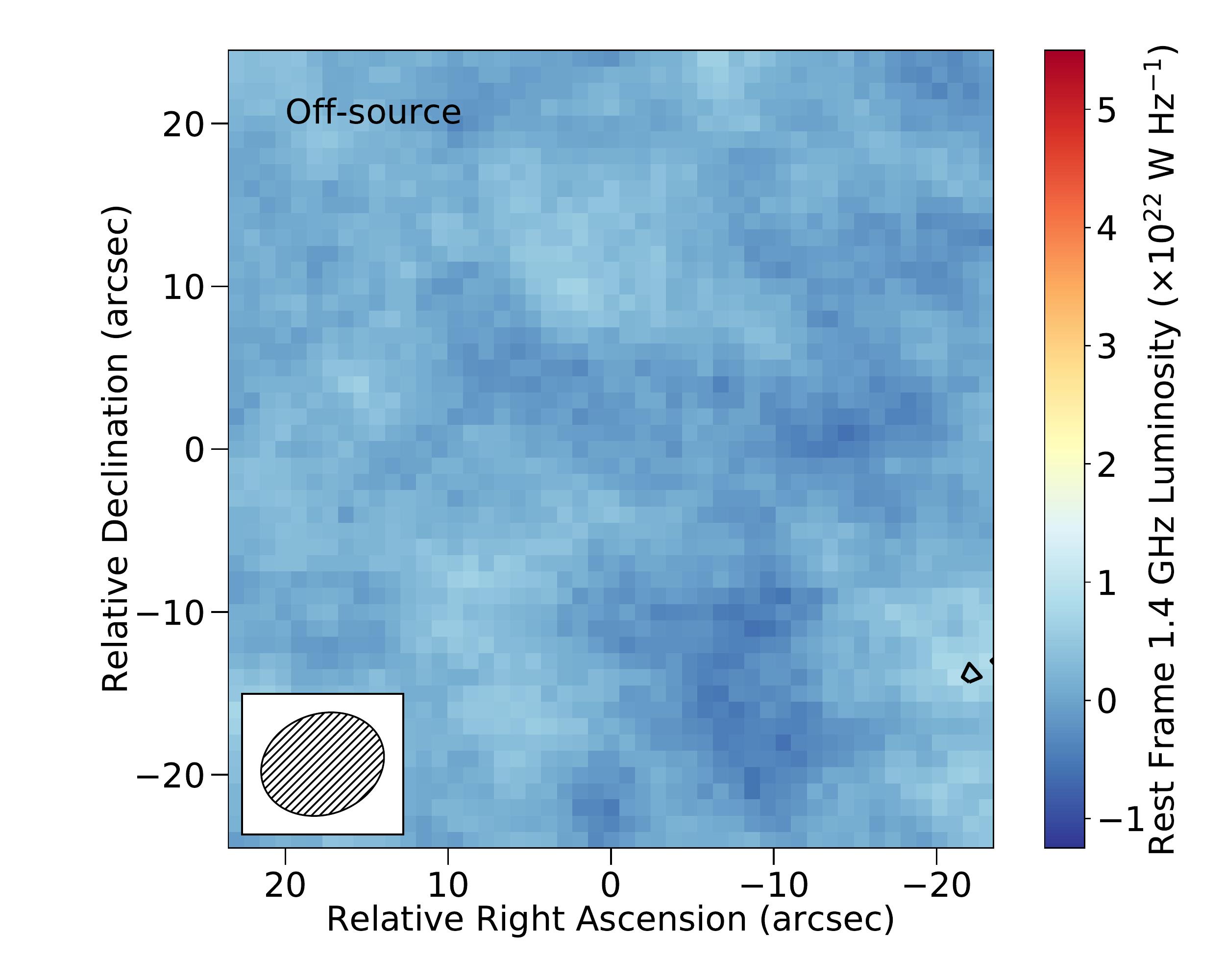}
\caption{The stacked rest-frame 1.4~GHz radio continuum luminosity (in units of 
$\rm 10^{22}\;W\;Hz^{-1}$) from [A, left panel]~the 3698 blue star-forming galaxies with 
$\rm M_B \leq -21$ of our sample, and [B, right panel]~locations offset by 
$100''$ from the 3698 DEEP2 galaxies. The RMS noise on each image is 
$\approx \rm 2.5 \times 10^{21}\;W\;Hz^{-1}$. A point source is clearly detected 
at the center of the left panel, while the right panel shows no evidence for emission. 
\label{fig:stack}}
\vskip 0.2in
\end{figure*}

\section{Stacking the radio continuum}
\label{sec:stacking}

We initially smoothed all four GMRT images with a uniform elliptical Gaussian beam of 
FWHM~$ = 6.1'' \times 4.8''$, which contains the synthesized beams of all the individual sub-fields. 
We then estimated the local RMS noise in a box of size $50'' \times 50''$ centered at the location 
of each DEEP2 galaxy, and used this to generate the RMS noise distribution for each field. 
This was used to exclude DEEP2 galaxies with high local RMS noise, in the upper 10\% tail of 
the RMS noise distribution for each field. Varying the exclusion threshold (between 5\% and 20\%) 
did not significantly affect our results. We also only included DEEP2 galaxies lying within the 
FWHM of the GMRT primary beam, to reduce the effect of deconvolution errors, which increase 
significantly below the half-power point of the primary beam.

It is important to exclude active galactic nuclei (AGNs) from the sample, in order to 
interpret the stacked radio emission as arising from star formation. Studies of galaxy radio 
luminosity functions have found that 1.4~GHz radio luminosities $\gtrsim 2 \times 10^{23}$~W~Hz$^{-1}$ 
arise mostly from AGNs, while lower luminosities are produced by star formation 
\citep[e.g.][]{sadler02,condon02,smolcic08}. The $5\sigma$ detection threshold in our images 
($\approx 70-195 \mu$Jy corresponds to a 1.4~GHz radio luminosity of $\approx (4-10) \times 
10^{23}$~W~Hz$^{-1}$ at the median redshift, $z \approx 1$, of our targets. All individual 
radio detections are hence likely to arise from AGNs (or from extreme starburst galaxies). To
reduce AGN contamination, we hence excluded from the stack any DEEP2 galaxy detected at 
$\geq 5\sigma$ significance with respect to its local RMS noise.

We focused on blue, star-forming galaxies, with ``color''~$\rm C \leq 0$, where 
$\rm C = (U-B) + 0.032 (M_B+21.62)-1.035$ \citep{willmer06}. Further, we only considered galaxies with 
$\rm M_B \leq -21$, as this yields a near-complete, absolute-magnitude-limited sample \citep{newman13}. 
However, we note that the DEEP2 survey is not strictly complete at $\rm M_B \leq -21$ over $0.7 \lesssim z \lesssim 1.45$, 
as there are galaxies in the survey with currently unknown redshifts \citep[e.g. redshift 
quality $<3$;][]{newman13}. 

The stacking was carried out in rest-frame 1.4~GHz luminosity $\rm L_{1.4\;GHz}$, rather than flux density, 
to correctly account for the different luminosity distances and redshifts of individual galaxies. This was 
done by shifting a $50'' \times 50''$ sub-image centered on each galaxy from flux density to 
$\rm L_{1.4\;GHz}$, assuming a spectral index of $\alpha = -0.8$ \citep[with flux density 
$S_\nu \propto \nu^\alpha$; e.g. ][]{condon92}, before carrying out the stacking procedure. 
We used ``median stacking'' as the median is more robust against outliers than the mean, and 
provides information on ``typical'' members of the target population \citep{white07}. 
The use of median stacking reduces the effect of contamination by undetected radio emission (e.g. 
lying just below our $5\sigma$ detection threshold) from individual AGNs or starburst galaxies. Finally,
the stacking was carried out using $50'' \times 50''$ sub-images, centered at the galaxy locations; the 
same procedure was used to stack regions $100''$ away from the galaxies, to test for systematic effects.


\section{Results and Discussion}
\label{sec:results}

\subsection{The total SFR of the DEEP2 galaxies}
\label{subsec:total_sfr}

Our sample contains 3698 blue star-forming galaxies with $\rm M_B \leq -21$, a median redshift 
$z_{\rm med} = 1.1$ and a median stellar mass $\rm M_\star = 10^{10.3} \; M_\odot$. Fig.~\ref{fig:stack}[A]
shows the median-stacked image of these galaxies: an unresolved source, detected at $\approx 17\sigma$ 
significance with $\rm L_{1.4 \; GHz} = (4.13 \pm 0.24) \times 10^{22}$~W~Hz$^{-1}$, is clearly 
visible. Fig.~\ref{fig:stack}[B] shows the image obtained from stacking locations offset by $100''$ from 
the DEEP2 galaxies; this shows no evidence of either emission or systematic effects.

To obtain the total SFR from the 1.4~GHz luminosity, we adopt the calibration of \citet{yun01},
\begin{equation}
\label{eqn:sfryun}
\rm SFR \;(M_\odot \; yr^{-1}) = (5.9 \pm 1.8) \times 10^{-22} \; L_{1.4\;GHz} ({W \; Hz}^{-1}) \;,
\end{equation}

which assumes a Salpeter initial mass function (IMF), with masses in the range $\rm (0.1-100) \; M_\odot$. The 
$\approx 30$\% uncertainty in this relation arises primarily from estimates of the local SFR density 
\citep{yun01}; this systematic uncertainty has not been included in our error estimates below. 
Using our measured rest-frame 1.4~GHz luminosity in equation~(\ref{eqn:sfryun}) yields a median SFR of 
$\rm SFR_{RADIO} = (24.4 \pm 1.4) ~M_\odot$~yr$^{-1}$ for the 3698 galaxies of our sample. 

The fact that the stacked radio emission is unresolved implies a transverse size $\lesssim 8$~kpc at 
$z_{\rm med} = 1.1$. Note that any uncorrected phase errors arising from the ionosphere would increase 
the observed spatial extent of the radio emission. Star formation in the DEEP2 galaxies thus appears to 
typically arise from the central regions, of size $\ll 8$~kpc.

\begin{figure*}[t]
\centering
\includegraphics[scale=0.79,trim={1cm 2cm 0.5cm 1cm},clip]{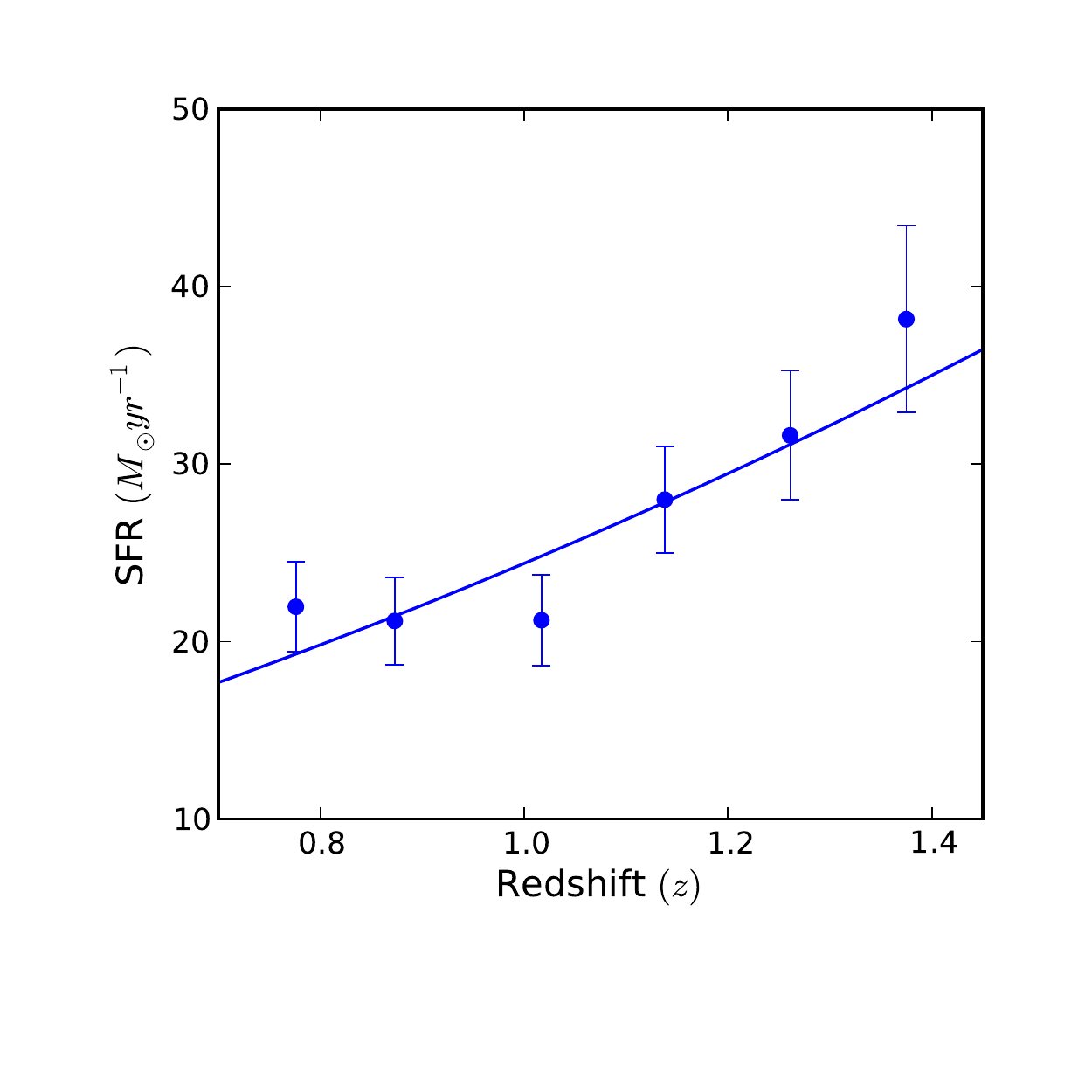}
\includegraphics[scale=0.79,trim={0.5cm 2cm 1.2cm 1cm},clip]{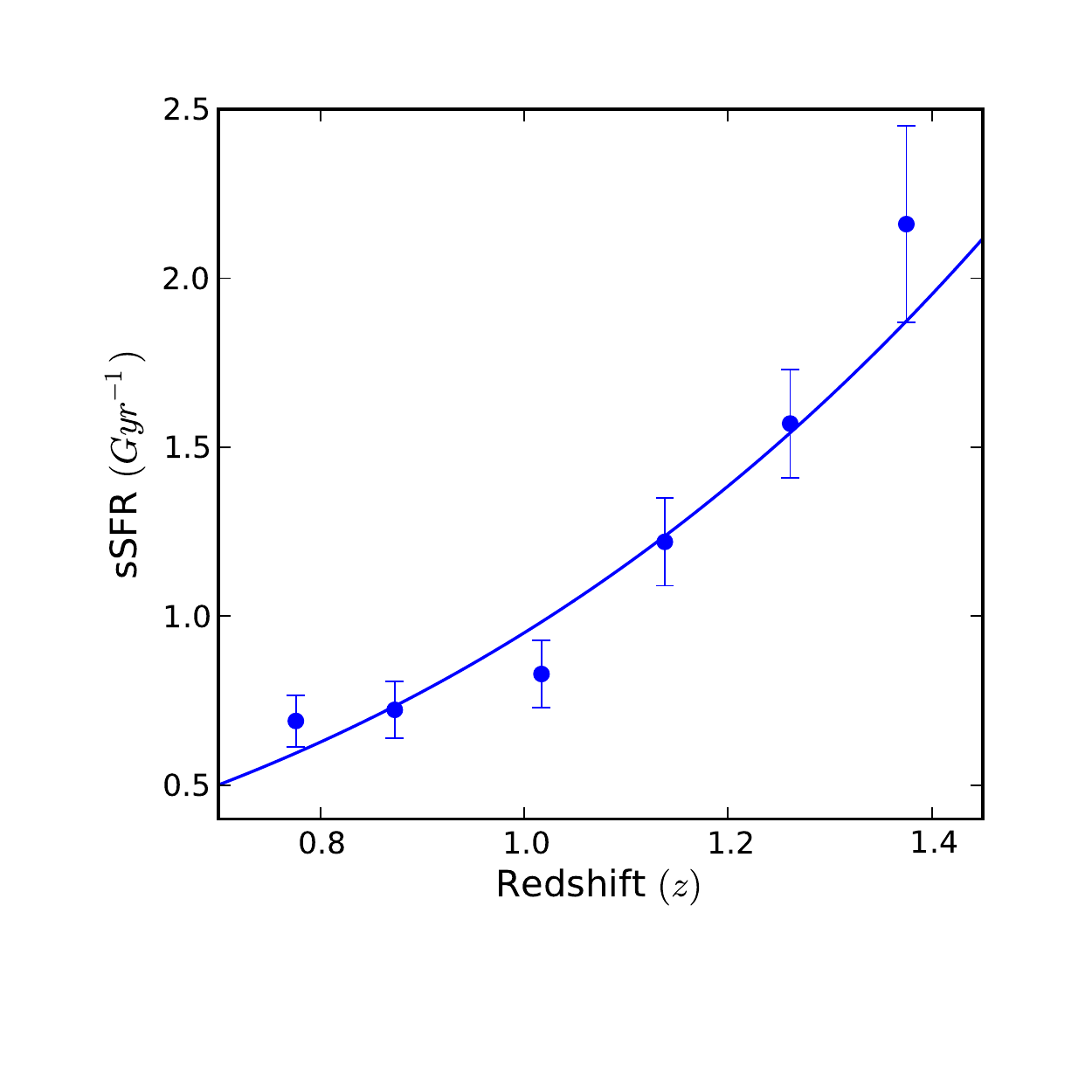}
\caption{Redshift evolution of [A]~the SFR (left panel) and [B]~the sSFR (right panel) for the 
3698 blue star-forming galaxies of our sample. The solid curves show the best-fit power-law models 
for the redshift evolution, with SFR~$\propto (1+z)^{1.98 \pm 0.50}$ and sSFR~$\propto (1+z)^{3.94 \pm 0.57}$.}
\label{fig:sfrz}
\end{figure*}

\begin{figure*}[t]
\centering
\includegraphics[scale=0.79,trim={1cm 2cm 0.5cm 1cm},clip]{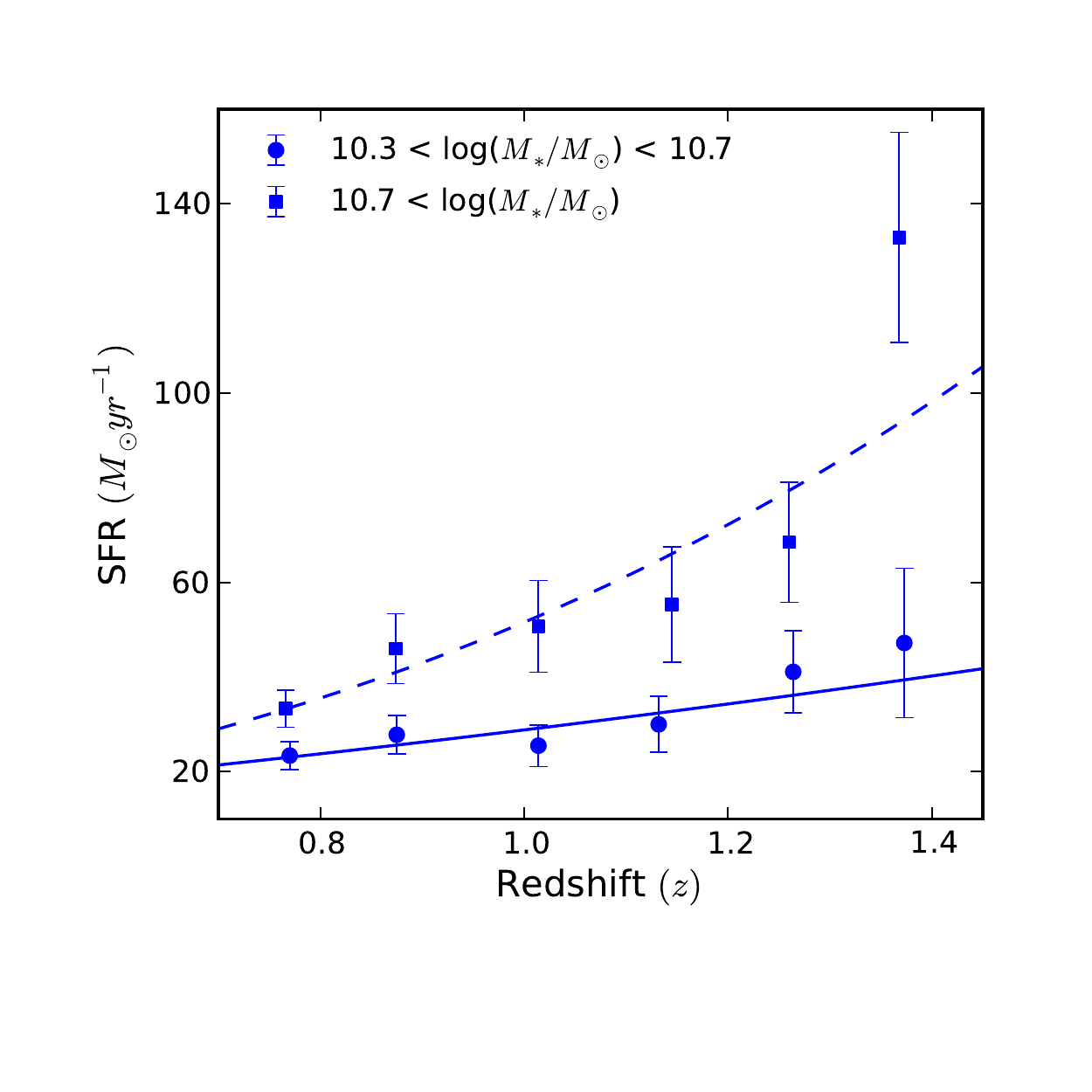}
\includegraphics[scale=0.79,trim={0.5cm 2cm 1.2cm 1cm},clip]{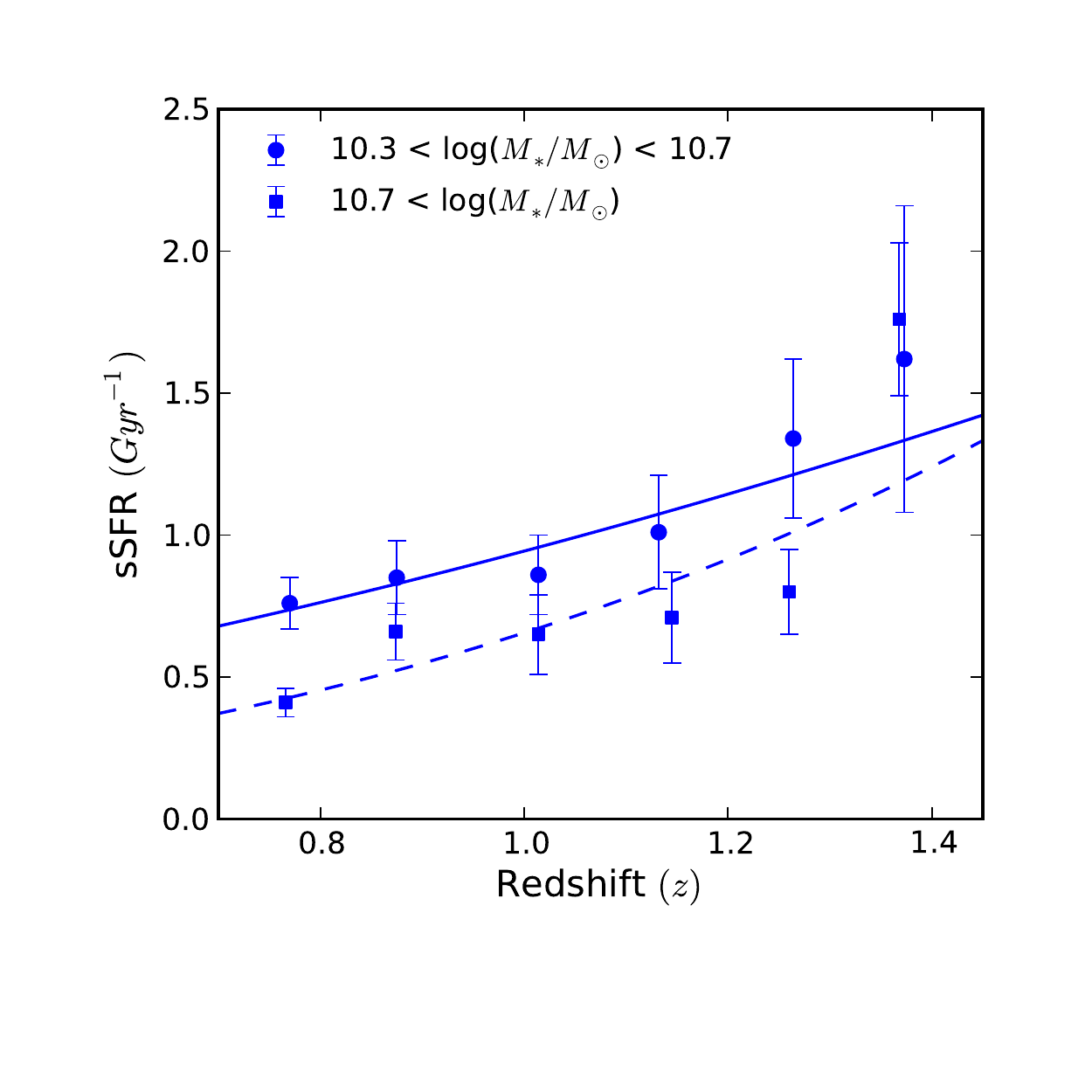}
\caption{Redshift evolution of [A]~the SFR (left panel) and [B]~the sSFR (right panel) for DEEP2 galaxies 
in two stellar mass ranges, $\rm 10.3 < log[M_\star/M_\odot] < 10.7$ and $\rm log[M_\star/M_\odot] > 10.7$.
For both mass ranges, the median stellar mass in the different redshift bins is approximately the same.
The solid and dashed curves show the best-fit power-law models for the two different mass ranges.
The power-law exponents for the SFR evolution are $1.83 \pm 0.53$ and $3.53 \pm 0.72$ for, respectively,
$10.3 \rm < log[M_\star/M_\odot] < 10.7$ and $\rm log[M_\star/M_\odot] > 10.7$, while the exponents for the 
sSFR evolution are $2.02 \pm 0.42$ and $3.44 \pm 0.94$, respectively.}
\label{fig:sfr_m_z}
\vskip 0.2in
\end{figure*}

\begin{figure*}[t]
\centering
\includegraphics[scale=0.79,trim={0.6cm 2cm 0.5cm 1cm},clip]{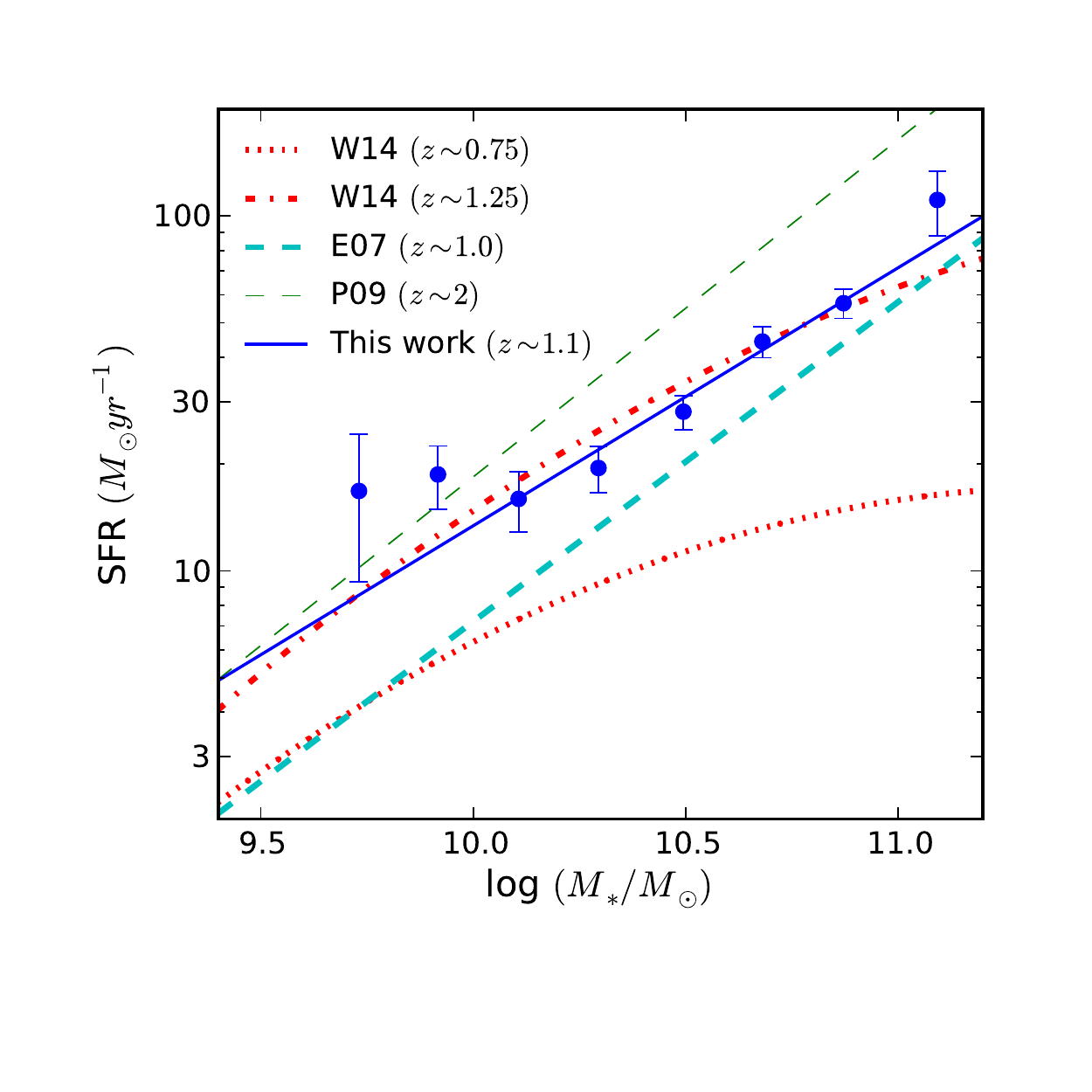}
\includegraphics[scale=0.79,trim={0.5cm 2cm 1.2cm 1cm},clip]{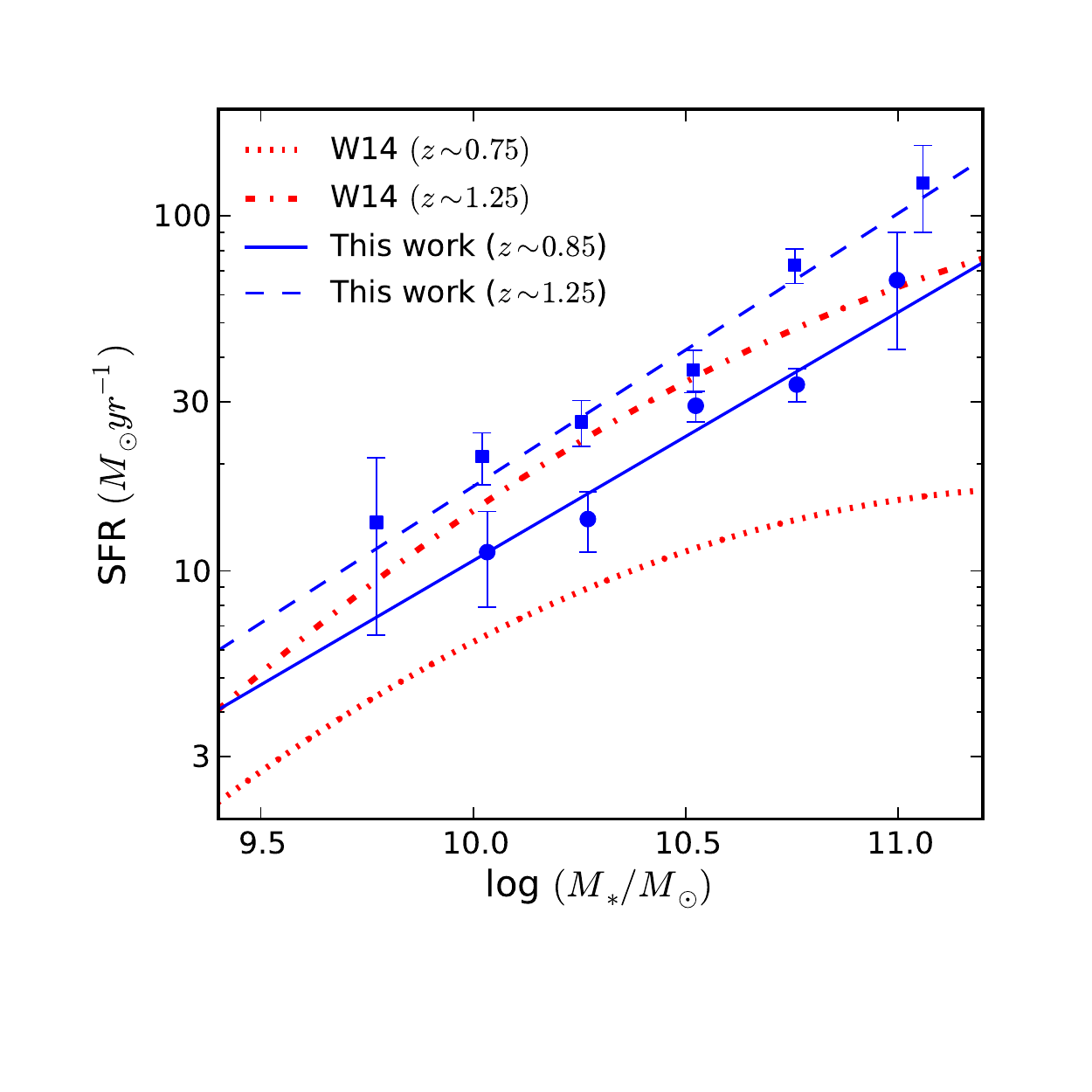}
\caption{[A]~The ``main sequence'', with median SFR plotted versus median stellar mass for the 3698 galaxies 
of our full sample; the best-fit main-sequence relation is shown by the solid blue line. The main 
sequence relations obtained by \citet[][E07; cyan dashed line, at $z \approx 1$]{elbaz07}, \citet[][P09; black 
dashed line, at $z \approx 2$]{pannella09}, and \citet[][W14; red dotted and dashed-dotted curves, for 
redshift bins centered at $z\approx 0.75$ and $z\approx 1.25$, respectively]{whitaker14} are also shown, 
for comparison. [B]~The redshift evolution of the main sequence, with the low-$z$ ($0.7 < z < 1.0$) 
sub-sample  indicated by circles, and the high-$z$ ($1.0 < z < 1.45$) sub-sample by squares. The 
solid and dashed blue lines indicate our main sequence relations for the low-$z$ and high-$z$ 
sub-samples, respectively, while the relations of \citet{whitaker14} at $z \approx 0.75$ and 
$z \approx 1.25$ are again shown as red dotted and dashed-dotted curves, respectively. See main 
text for discussion.}
\label{fig:sfrsm}
\vskip 0.2in
\end{figure*}

\subsection{SFR evolution with redshift}
\label{subsec:sfr_z}

We examined the dependence of the SFR and the specific star formation rate (sSFR, defined as the SFR per 
unit stellar mass) on redshift by dividing the sample into six uniformly-spaced redshift bins 
covering $z \approx 0.7-1.45$, and carrying out the median stacking independently for the galaxies in each bin. 
For the sSFR, the stacking was carried out in the ratio $\rm L_{1.4\;GHz}/M_\star$ for each galaxy, 
so as to not introduce errors from averaging in the stellar mass. Figs.~\ref{fig:sfrz}[A] and [B] 
show, respectively, the median SFR and the median sSFR of each sub-sample plotted against the median 
redshift of its bin. We find that both the SFR and the sSFR decrease with decreasing redshift, with 
SFR~$\propto (1+z)^{1.98 \pm 0.50}$ and sSFR~$\propto (1+z)^{3.94 \pm 0.57}$ over $0.7\lesssim z \lesssim 1.45$. This 
is consistent with earlier studies, mostly based on optical SFR indicators, which have obtained 
sSFR~$\propto (1+z)^\beta$ with $\beta \approx 2.8-3.8$ for $0 < z \lesssim 2.5$ 
\citep[e.g.][]{karim11,fumagalli14,tasca15,ilbert15}. We note that median SFR of our 
galaxies increases less steeply with redshift than the median sSFR because the median stellar mass 
of the sample decreases with increasing redshift \citep[see also][for \textit{Herschel}-detected 
galaxies at $1.2 < z < 4$]{sklias17}. 

To study the SFR evolution of galaxies with similar stellar mass, we chose sub-samples of the DEEP2 
galaxies in two different mass ranges, $\rm 10.3 <log[M_\star/M_\odot]< 10.7$ and $\rm log[M_\star/M_\odot]> 10.7$. 
The mass ranges were chosen to ensure that the median stellar mass of each sub-sample is approximately 
equal in each redshift bin. Figs.~\ref{fig:sfr_m_z}[A] and [B] show the redshift evolution of SFR and sSFR, 
respectively, for galaxies in the above mass ranges. We find that the SFR increases with redshift as 
SFR~$\propto (1+z)^{1.83 \pm 0.53}$ and SFR~$\propto (1+z)^{3.53 \pm 0.72}$ for, respectively, the lower-mass 
($\rm 10.3 <log[M_\star/M_\odot]< 10.7$) and higher-mass ($\rm log[M_\star/M_\odot]> 10.7$) sub-samples,
while the sSFR increases $\propto (1+z)^{2.02 \pm 0.42}$ (low-mass sub-sample) and 
$\propto (1+z)^{3.49 \pm 0.94}$ (high-mass sub-sample). We thus find only marginal ($\approx 2 \sigma$ 
significance) evidence that the SFR increases more steeply with redshift for the higher-mass sub-sample in our 
data. Indeed, excluding the last redshift bin from our fits reduces the statistical significance of the 
difference in the exponents of the two sub-samples further, to $\approx 1 \sigma$ significance. We thus 
find no statistically significant evidence of a difference in the redshift evolution of the SFR for low-mass 
and high-mass galaxies, with stellar masses $\rm log[M_\star/M_\odot] > 10.3$ in the redshift range 
$0.7  \lesssim  z  \lesssim  1.45$. 

We note, in passing, that the DEEP2 galaxy sample is not complete in stellar mass. As such, it is 
possible that massive dusty galaxies with high SFRs have been excluded from our stellar mass sub-samples 
due to dust obscuration. This effect is likely to vary with redshift as the observed optical bands 
translate to shorter rest-frame UV wavelengths at higher redshifts, with correspondingly higher dust 
extinction effects. Any dependence of the mass completeness of the sample on redshift, if present 
in our sample, is likely to affect our results.

\begin{figure*}[t]
\centering
\includegraphics[scale=0.58,trim={1.1cm 2cm 0.8cm 1cm},clip]{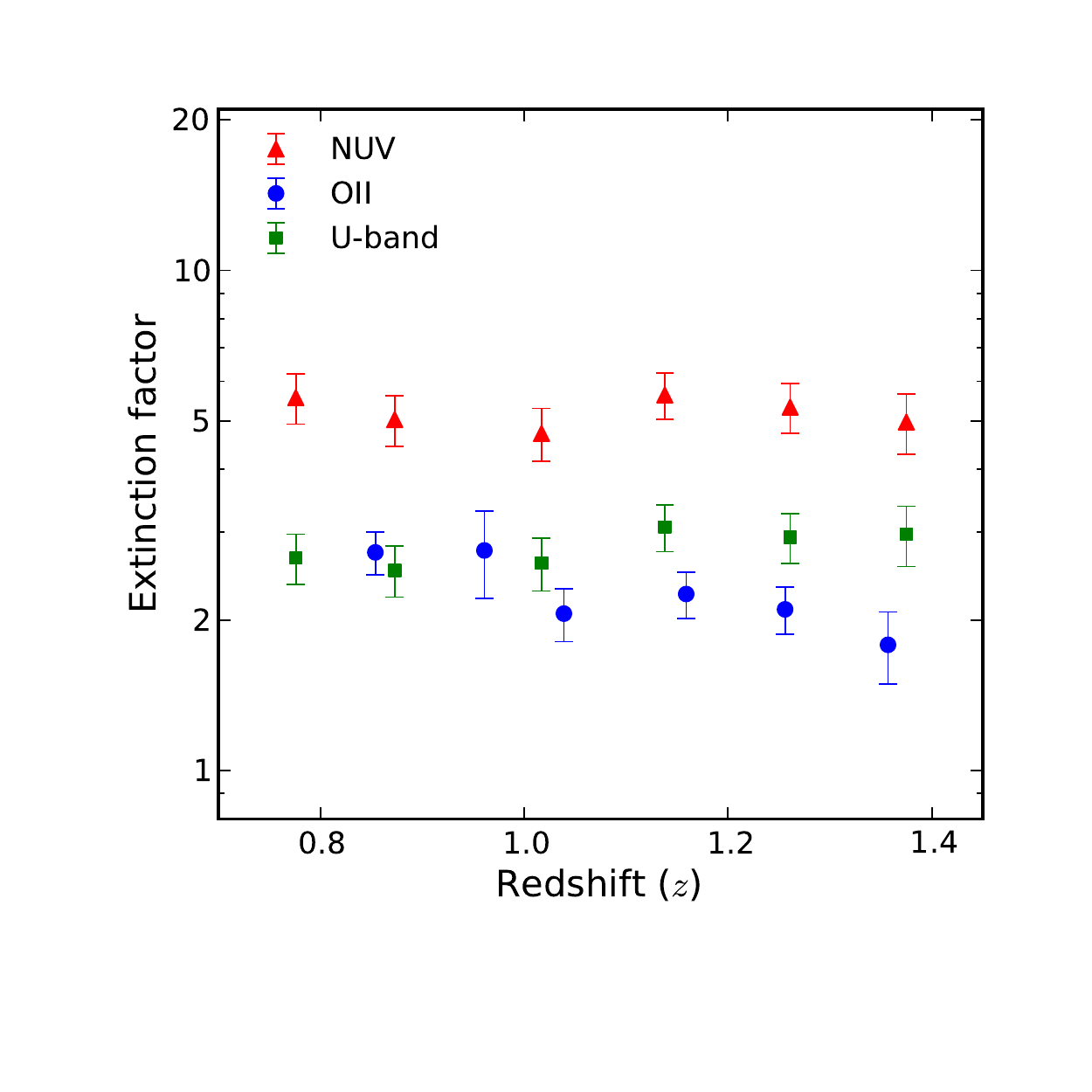}
\includegraphics[scale=0.58,trim={1.9cm 2cm 0.6cm 1cm},clip]{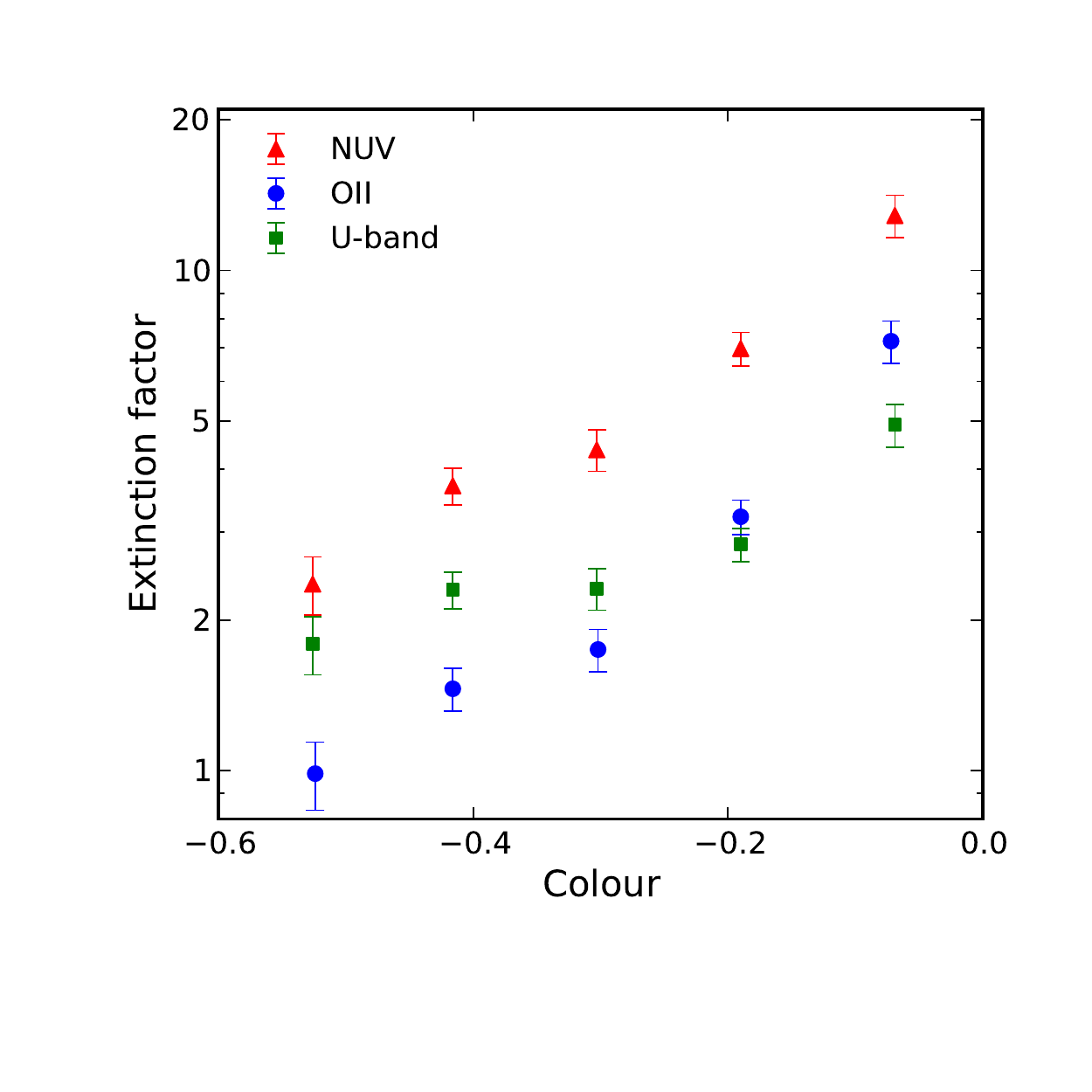}
\includegraphics[scale=0.58,trim={1.9cm 2cm 1.2cm 1cm},clip]{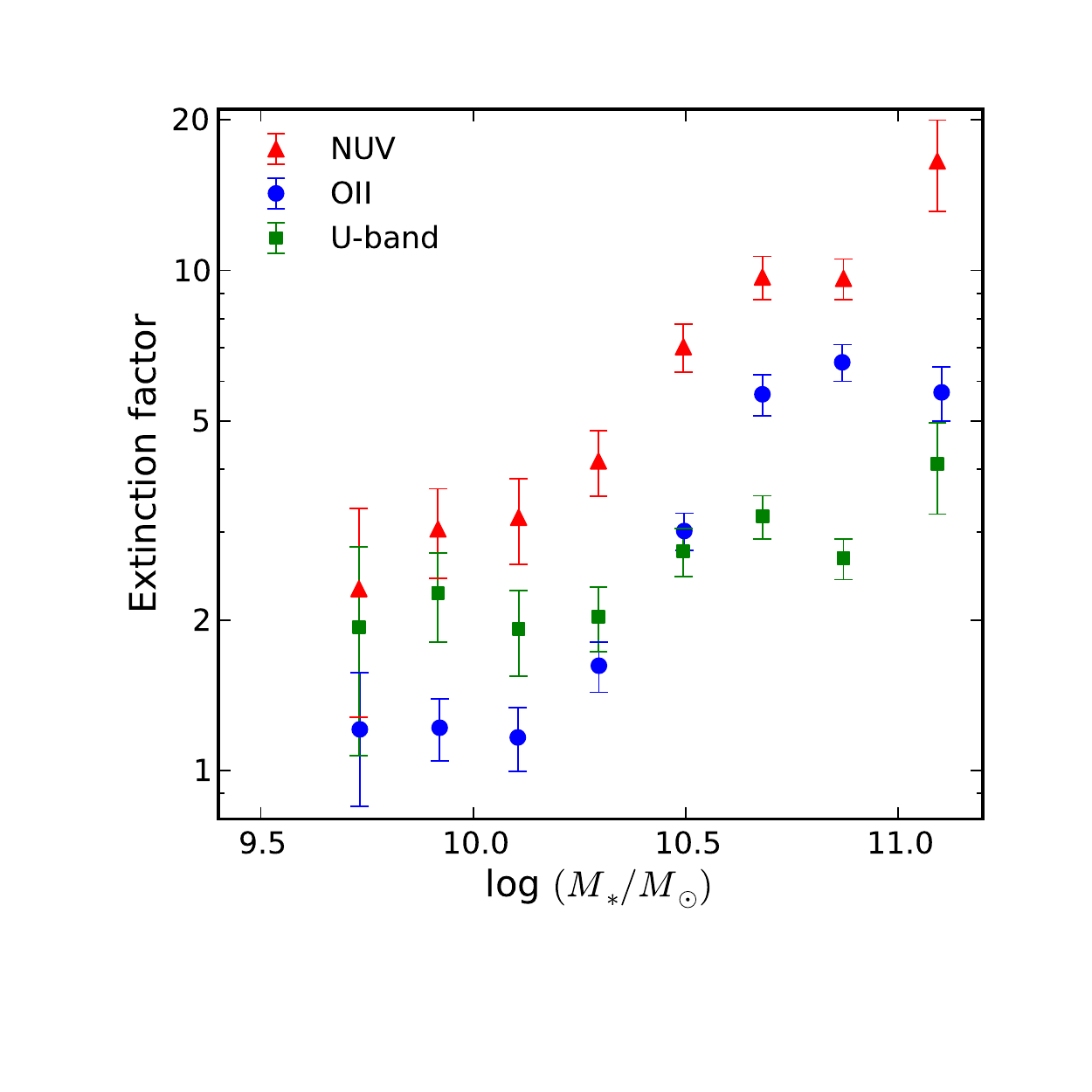}
\vskip 0.01in
\caption{The dust extinction correction factors for the rest-frame 230~nm NUV continuum 
luminosity (triangles), the rest-frame U-band continuum luminosity (squares), and the 
[O{\sc ii}]$\lambda$3727 line luminosity (circles) plotted against [A, left panel]~redshift, 
[B, middle panel]~galaxy color, and [C, right panel]~stellar mass, for the 3698 DEEP2 
galaxies in our sample with $\rm M_B \leq -21$. See main text for discussion.
\label{fig:extfac}}

\end{figure*}

\subsection{The main sequence of star-forming galaxies}
\label{subsec:ms}

Next, we studied the dependence of the SFR on $\rm M_\star$ by stacking galaxy sub-samples binned by 
stellar mass. Fig.~\ref{fig:sfrsm}[A] shows the median SFR of each sub-sample plotted against its 
median $\rm M_\star$: a tight correlation between SFR and $\rm M_\star$ \citep[usually referred to 
as the ``main sequence''; e.g. ][]{noeske07} is clear in the figure, with SFR~$\rm = (13.4 \pm 1.8) \times 
\left[M_\star/(10^{10} \; M_\odot) \right]^{(0.73 \pm 0.09)}$~M$_\odot$~yr$^{-1}$. Our 
main-sequence power-law index of $\alpha \approx 0.73 \pm 0.09$ is in good agreement with earlier 
studies at similar redshifts ($z \approx 0.5-2.5$), which have typically obtained $\alpha \approx 0.6-0.9$ 
\citep[e.g.][]{noeske07,elbaz07,santini09,pannella15}. 

\citet{whitaker14} studied the main sequence relation for mass-complete samples at different redshift 
intervals in the range $0.5 < z < 2.5$, finding evidence for a flattening of the slope of the 
main sequence at high stellar masses in all redshift bins. Our main sequence relation at 
$z_{med}=1.1$ is in good agreement with the relation obtained by \citet{whitaker14} at 
$z \approx 1.25$. However, our sensitivity is too low to confirm the presence of a changing slope in 
the main sequence.

We examined the evolution of the main sequence with redshift by dividing the full sample into two 
redshift bins ($0.7 < z < 1.0$ and $1.0 < z < 1.45$), and then independently applying the median stacking procedure 
to six stellar mass bins within each redshift bin. Fig.~\ref{fig:sfrsm}[B] shows the median SFR plotted 
against the median $\rm M_\star$ for galaxies in the high-$z$ and low-$z$ sub-samples, indicated by filled squares
and circles, respectively. We obtained SFR~$\rm = (10.7 \pm 2.3) \times \left[M_\star/(10^{10} \; 
M_\odot)\right]^{(0.70 \pm 0.15)}$~M$_\odot$~yr$^{-1}$ for the low-$z$ sub-sample ($z_{\rm med} \approx 0.85$) and 
SFR~$\rm = (17.3 \pm 2.1) \times \left[{\rm M}_\star/(10^{10} \; M_\odot)\right]^{(0.77 \pm 0.09)}$~M$_\odot$~yr$^{-1}$ 
for the high-$z$ sub-sample ($z_{\rm med} \approx 1.20$). We thus find only weak ($\approx 2\sigma$ significance) 
evidence that the normalization of the main sequence increases with increasing redshift, by a factor of 
$\approx 1.6$ between $z \approx 0.85$ and $z \approx 1.2$. Similar results have been obtained in 
earlier studies, albeit mostly using either optical SFR indicators \citep[e.g.][]{noeske07}, or galaxy samples with photometric 
redshifts \citep[e.g.][]{santini09,pannella15}. However, we find no evidence that the slope of the main 
sequence evolves with redshift, over $0.7 \lesssim z \lesssim  1.45$, contrary to the expected decline in slope 
with time expected in cosmic downsizing scenarios \citep{cowie96}, due to the shifting of star formation activity
from more-massive to less-massive galaxies with time \citep[see also][]{santini09,pannella15}. 

Our main sequence relation at $z\approx 1.25$ is consistent within the errors with the relation obtained 
by \citet{whitaker14} at the same redshift. However, our relation at $z\approx 0.85$ is significantly 
above that of \citet{whitaker14} at $z\approx 0.75$. This is consistent with a decline in the normalization 
of the main sequence with decreasing redshift \citep[e.g.][]{elbaz07}, as the two low-$z$ sub-samples have 
different redshift distributions, $0.5 < z < 1.0$ for \citet{whitaker14} and $0.7 <z < 1.0$ for this work. 

However, we again note that the DEEP2 galaxies are selected to have $R < 24.1$, i.e. are selected 
at rest-frame wavelengths of $\approx 2700-3800$~\AA, and that the sample is not mass-complete. Massive 
dusty galaxies with high SFRs are likely to have been excluded due to their high extinction. Our ``color"
selection also excludes some of these galaxies as these are likely to be red in color. This might
affect our results by flattening the slope of the main sequence. The effect is likely to be 
stronger for higher-redshift galaxies, as these are selected at rest-frame UV wavelengths, $< 3000$~\AA, 
where the extinction effects are larger. This could cause our inferred main-sequence slopes to appear flatter, 
especially in the high-$z$ sub-sample.

\subsection{Comparisons with SFRs from other tracers: Dust extinction}
\label{subsec:extinction}

The SFRs of high-$z$ star-forming galaxies, such as the DEEP2 galaxies of our sample, are usually 
estimated from optical or near-IR imaging or spectroscopy, based on SFR tracers such as the rest-frame 
UV continuum luminosity or the [O{\sc ii}]$\lambda$3727 or H$\alpha$ line luminosity.  However, the 
measured luminosities in these SFR tracers are reduced from their intrinsic values due to extinction by 
dust in the interstellar medium of the target galaxy. The use of such tracers hence usually under-estimates 
the SFR in high-$z$ galaxies. A variety of prescriptions are available in the literature to correct for the 
dust attenuation \citep[e.g.][]{salim07,hao11}, but their accuracy is unknown, as dust extinction effects 
are likely to depend on the physical properties of the galaxies (e.g. stellar mass, color, metallicity, etc). 
In the case of the DEEP2 galaxy sample, we have estimated the median {\it total} SFR (i.e. both 
unobscured and obscured) from the rest-frame 1.4~GHz radio continuum, which is unaffected by dust extinction. 
In this section, we estimate the SFRs of different sub-samples of the DEEP2 galaxies from a set of optical/UV 
tracers that are commonly used for high-$z$ galaxies (the near-ultraviolet (NUV) continuum luminosity, the 
rest-frame U-band continuum luminosity, and the [O{\sc ii}]$\lambda$3727 line luminosity). We then compare 
these median SFR estimates to the radio-derived median SFRs to infer the correction for dust extinction that 
should be applied to the SFR inferred from each tracer for similar populations of star-forming galaxies, as 
a function of color, stellar mass and redshift. The SFR correction factor will be referred to as the dust 
extinction correction factor $\epsilon$, defined as 
\begin{equation}
\rm \epsilon_X \equiv SFR_{RADIO}/SFR_X \;,
\label{eq:ext_def}
\end{equation}
where X corresponds to the SFR tracer in question (i.e. the NUV continuum luminosity, the U-band continuum 
luminosity, the [O{\sc ii}]$\lambda$3727 line luminosity, etc), and $\epsilon_X \geq 1$.

The [O{\sc ii}]$\lambda$3727\AA\ line luminosity is a popular SFR tracer in high-$z$ galaxies, especially 
for objects at $z \approx 0.7-2.5$, for which the H$\alpha$ line is redshifted to near-IR wavelengths. 
In the case of the DEEP2 galaxies of our sample, reliable [O{\sc ii}]$\lambda$3727\AA\ line luminosities 
are available for galaxies at $0.8 < z < 1.4$, and one can hence immediately infer the SFR for these 
galaxies from this tracer, using standard calibrations \citep[e.g.][]{kennicutt98}. Following \citet{weiner07}, 
we assume a line ratio [O{\sc ii}]$\lambda$3727/H$_{\alpha} =0.69$, appropriate for high-redshift galaxies.
We then apply the calibration \citep[valid for a Salpeter IMF; ][]{kennicutt98}

\begin{equation}
\label{eqn:oiisfr}
{\rm [SFR_{OII}/ M_\odot \; yr^{-1}]} = 7.9 \times 10^{-42} {\rm [L_{H\alpha}/ergs\; s^{-1}]} \;,
\end{equation}  

to obtain a median SFR of $\rm SFR_{OII} = 12.5 \; M_\odot$~yr$^{-1}$. Combining this with our 
median radio SFR estimate yields a median dust extinction correction factor of 
$\rm \epsilon_{OII} \approx 2.0$ for the [O{\sc ii}]$\lambda$3727 SFR estimator. We emphasize that this
correction factor is applicable for main-sequence, star-forming galaxies with $\rm M_B \leq -21$.

Next, the observed B-band magnitudes of our 3,698 target galaxies are available from the DEEP2 survey. 
Since the sample galaxies lie in the redshift range $0.7 \lesssim z \lesssim 1.45$, the observed-frame 
B-band corresponds to rest-frame emission wavelengths $180 - 260$~nm, i.e. at NUV wavelengths. About 90\% 
of the NUV continuum emission of a galaxy is provided by young stars \citep[of age below 200~Myr;][]{kennicutt12}. 
The rest-frame 230~nm NUV luminosity of a galaxy can hence be used to infer its SFR 
\citep[e.g.][]{murphy11,hao11,kennicutt12}, via the relation \citep[applicable for a Kroupa IMF;][]{kroupa03}
\begin{equation} 
\label{equ:nuvsfr}
{\rm log \left(\frac{SFR_{NUV}}{M_\odot \; yr^{-1}}\right)} = {\rm log \left(\frac{\nu L_{\nu}}{ergs \; s^{-1}}\right) - 43.17}\;
\end{equation} 

Assuming that this calibration applies to the rest-frame wavelength range of $180-260$~nm (note that 
the calibration is in $\nu \times L_{\nu}$ and hence the above assumption is good to first order), we 
apply it to the observed-frame B-band luminosities, to obtain a median SFR (after adding 0.15~dex to shift 
to a Salpeter IMF) of $\rm SFR_{NUV} = 5.2 \; M_\odot$~yr$^{-1}$. Comparing our median radio SFR with this 
median NUV SFR then yields $\rm \epsilon_{NUV} \approx 4.7$, for star-forming galaxies with 
$\rm M_B \leq -21$.

The DEEP2 survey also provides the rest-frame U-band ($\rm \lambda \approx 3600$ \AA) absolute magnitude 
of our target galaxies \citep{newman13}. The rest frame U-band continuum luminosity can be used to estimate 
the SFR following the prescription of \citet{hopkins03}, 
\begin{equation} 
\label{equ:ubandsfr}
{\rm log \left(\frac{SFR_U}{M_\odot \; yr^{-1}}\right)} = {\rm 1.186 \times log \left(\frac{L_U}{1.81 \times 10^{21}\; W\; Hz^{-1}}\right)}\;
\end{equation} 
which assumes a Salpeter IMF. This yields a median SFR of $\rm SFR_U = 9.6 \; M_\odot$~yr$^{-1}$.
Combining this U-band median SFR estimate with our median radio SFR then yields a dust extinction 
correction factor of $\rm \epsilon_{U} \approx 2.5$ for the U-band SFR, again for star-forming 
galaxies with $\rm M_B \leq -21$.

We thus find that the largest dust extinction correction factor is needed for the NUV 230~nm 
continuum luminosity, with $\epsilon_{\rm NUV} \approx 4.7$; the dust correction factors are 
similar for the rest-frame U-band continuum luminosity ($\rm \epsilon_U \approx 2.5$) and 
for the [O{\sc ii}]$\lambda$3727 line luminosity ($\rm \epsilon_{OII} \approx 2.0$).

We also examined the dependence of the dust extinction correction factor for the NUV-, U-band-
and O{\sc ii}-based SFR estimators on redshift, absolute B-band magnitude, color, and stellar mass. 
To examine the dependence of $\rm \epsilon_X$ on absolute B-band magnitude, we compared 
the dust extinction correction factors for galaxies with $\rm M_B \leq -20$ with the above estimates 
of $\rm \epsilon_X$ for $\rm M_B \leq  -21$. To do this, we restricted to the redshift range $0.7 < z < 1.0$,
for which the DEEP2 sample is complete down to $\rm M_B \leq  -20$ \citep{newman13}. Including all DEEP2 
galaxies with $\rm M_B\leq -20$ and restricting the redshift range to $0.7<z<1.0$, we obtained dust extinction 
correction factors of $\rm \epsilon_{NUV} \approx 3.9$, $\rm \epsilon_U \approx 2.5$, 
and $\rm \epsilon_{OII} \approx 1.6$. In the same redshift range, for $\rm M_B \leq -21$, we obtain 
$\rm \epsilon_{NUV} \approx 5.0$, $\rm \epsilon_U \approx 2.4$, $\rm \epsilon_{OII} \approx 2.2$. The 
dust extinction correction factor thus appears to be systematically larger for brighter galaxies, 
except for the calibration based on the U-band luminosity.

Fig.~\ref{fig:extfac}[A] shows the dependence of $\epsilon_{\rm X}$ on redshift for the three 
SFR tracers. We find no evidence that the dust extinction correction factor varies with redshift 
over $0.7<z<1.45$, for galaxies with $\rm M_B\leq -21$. Figs.~\ref{fig:extfac}[B] and [C] plot 
$\rm \epsilon_X$ against galaxy color and stellar mass, 
respectively, for the three SFR tracers. $\rm \epsilon_X$ is seen to increase with both increasing 
color (i.e. from bluer to redder galaxies) and increasing stellar mass for all three estimators, 
with a strong dependence on color and stellar mass for the NUV luminosity and the [O{\sc ii}]$\lambda$3727 
line luminosity, and a weaker dependence for the rest-frame U-band luminosity. This is 
unsurprising for the NUV emission, given that dust attenuation is most effective at the shorter 
wavelengths. The [O{\sc ii}]$\lambda$3727 line emission suffers relatively little dust extinction 
in bluer and less massive galaxies, but is strongly affected by attenuation effects in redder and 
more massive galaxies, making it a good tracer of the total SFR for blue galaxies with 
$\rm M_\star \leq  10^{10.4} \; M_\odot$, but a less reliable tracer for more massive galaxies. 

Finally, the dust extinction correction factor for the SFR estimate from the rest-frame U-band luminosity 
shows a relatively weak dependence on color and stellar mass, with $\rm \epsilon_U$ varying by less 
than a factor of $\approx 2$ over our color and stellar mass range. Further, we find no evidence 
that $\rm \epsilon_U$ varies with redshift (over $0.7 < z < 1.45$) or B-band absolute magnitude 
(comparing galaxies with $\rm M_B \leq -20$ with those with $\rm M_B \leq -21$, in the same redshift
range). After applying an average dust extinction correction factor of $\rm \epsilon_U \approx 2.5$, 
the rest-frame U-band luminosity thus appears to yield a reasonable tracer of the total SFR for 
blue, star-forming, main-sequence galaxies at $z \approx 0.7-1.5$. 

We emphasize that the U-band luminosity of a galaxy is only likely to be a good tracer of the SFR for 
galaxies dominated by young stellar populations \citep[e.g.][]{cram98,hopkins03}. Moreover, the 
U-band luminosity is known to depend on both the evolutionary timescale and the star formation 
history \citep[e.g.][]{bell03,hopkins03}, yielding a non-linear relation between the SFR and 
the U-band luminosity \citep{hopkins03}. Specifically, the U-band luminosity may be contaminated 
by emission from older stellar populations, especially in low-luminosity galaxies. As such, 
caution should be used when inferring the SFR from the rest-frame U-band luminosity. Despite
these caveats, the fact that the dust extinction correction factor for the SFR estimate from the 
U-band luminosity does not vary significantly with color, stellar mass, B-band absolute magnitude, 
or redshift for the sub-sample of DEEP2 galaxies with $\rm M_B \leq -21$, indicates that the 
rest-frame U-band luminosity may be an interesting tracer of the SFR in similar galaxies at $z \approx 1$.

\subsection{Nebular and stellar extinction}
\label{subsec:nebular}

\begin{figure*}[t]
\centering
\includegraphics[scale=0.53,trim={0.3cm 0cm 0.7cm 0.5cm},clip]{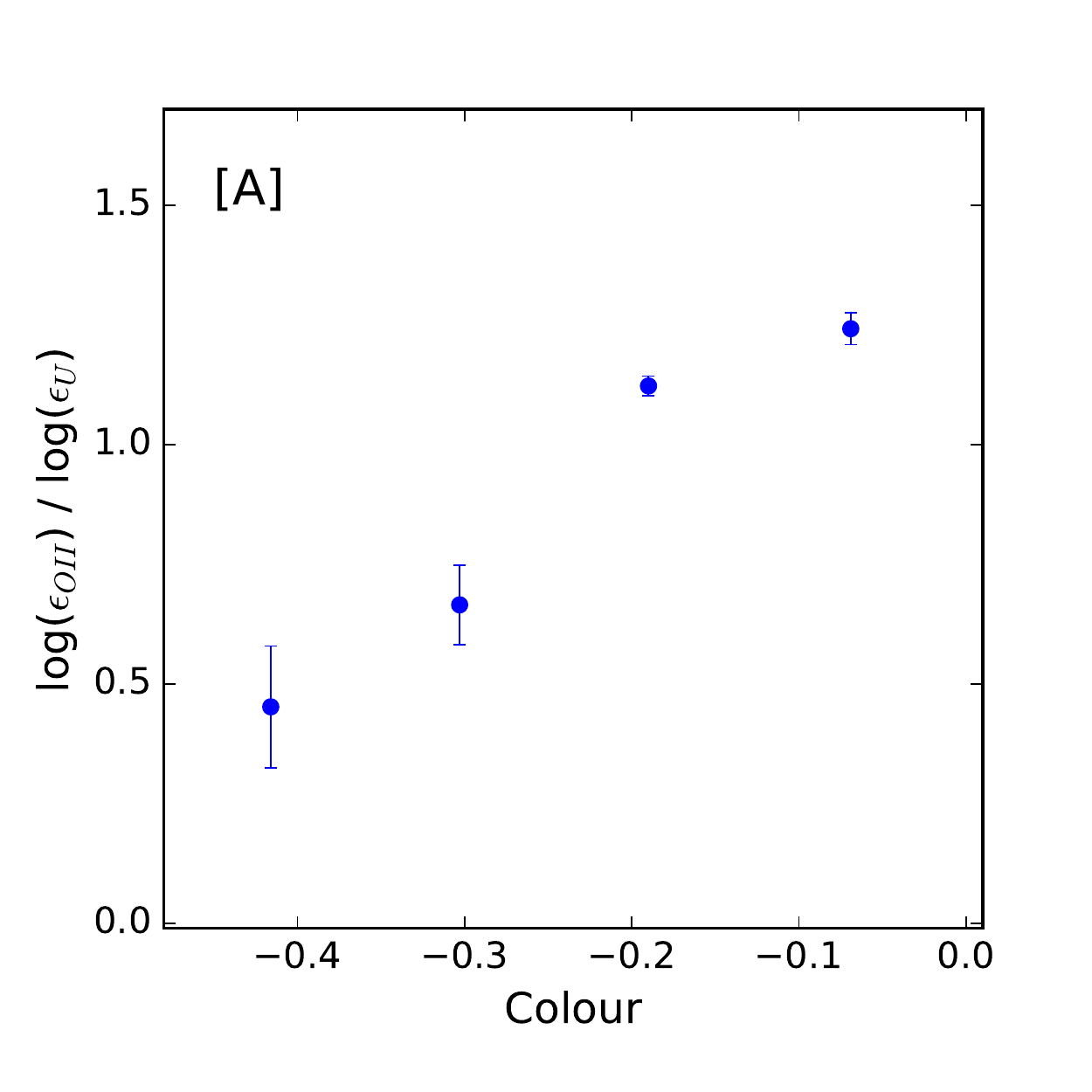}
\includegraphics[scale=0.53,trim={1.0cm 0cm 0.7cm 0.5cm},clip]{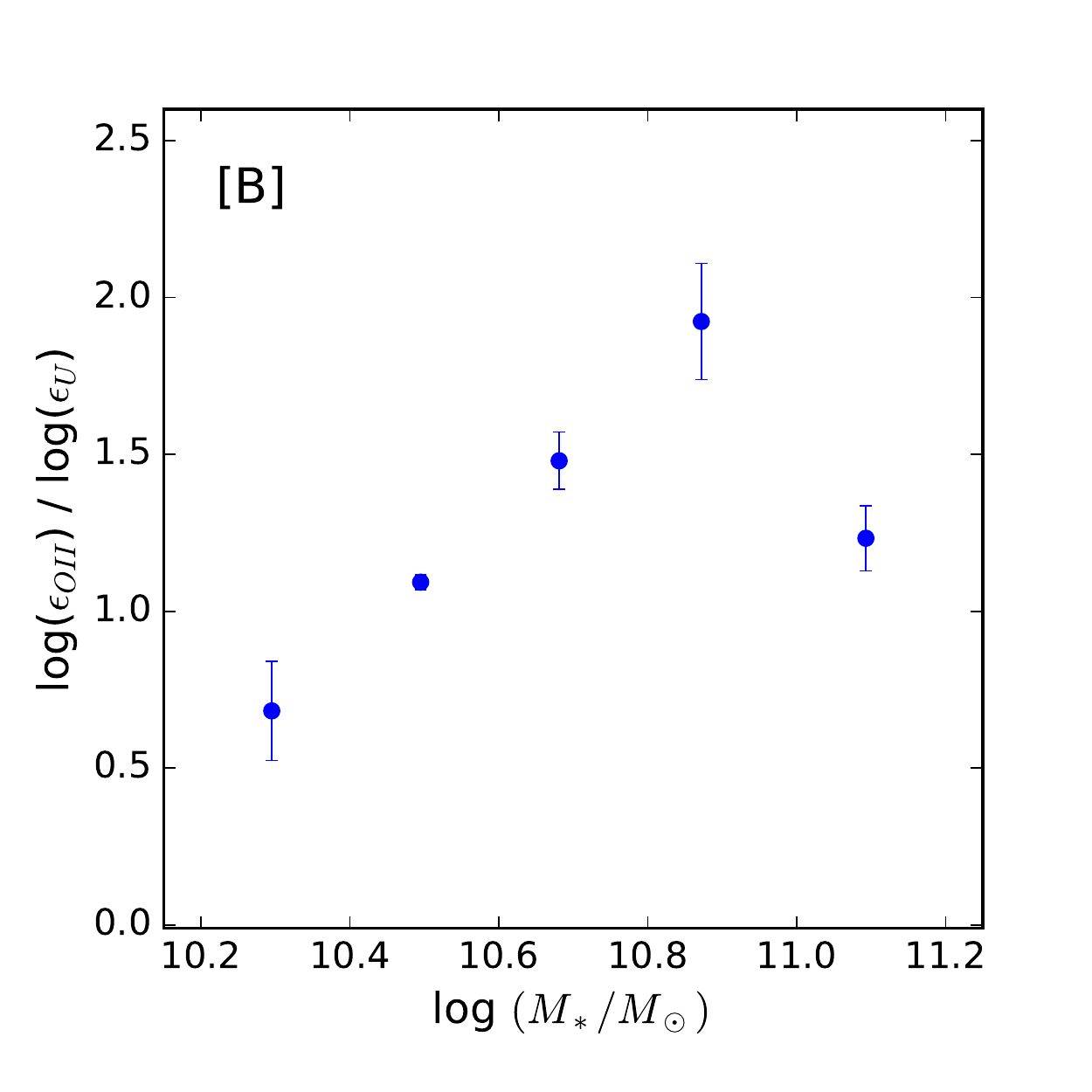}
\includegraphics[scale=0.53,trim={1.0cm 0cm 0.7cm 0.5cm},clip]{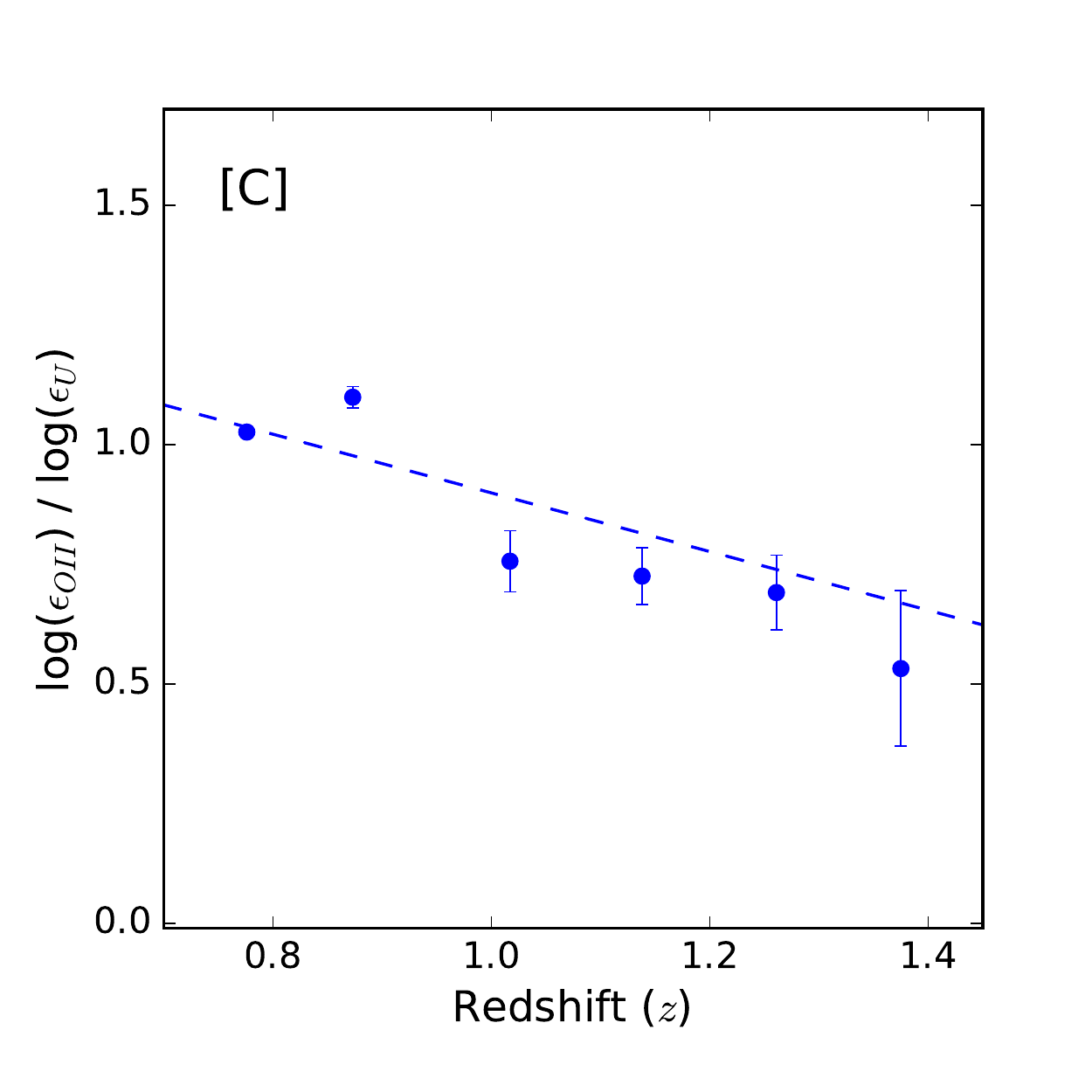}
\vskip 0.01in
	\caption{The ratio of the extinction factor for the O{\sc ii}$\lambda$3727\AA\ line to that 
	of the U-band continuum ($\rm E_{OII/U}$) plotted against [A]~galaxy colour, [B]~stellar mass, 
	and [C]~redshift,
	for the DEEP2 galaxies with $\rm M_B \leq -21$. The dashed line in the third panel shows 
	the linear fit $\rm E_{OII/U} = (1.51\pm0.24) - (0.61\pm0.30)\times z$. See main text for discussion.
\label{fig:o2u}}
\end{figure*}

In the local Universe, nebular emission lines have been shown to suffer more extinction, by a factor of $\approx 1.7$, 
than the stellar continuum at the same wavelength in a wide range of galaxy types
\citep[e.g.][]{fanelli88,calzetti94,calzetti97,mayya96,sullivan01,cidfernandes05,wild11}. This differential
extinction has been shown to depend on galaxy properties such as the SFR, the sSFR, the inclination, etc 
\citep[e.g.][]{sullivan01,cidfernandes05,wild11,battisti16,battisti17}. A two-component dust model, including a 
diffuse, optically-thin component arising from the galactic interstellar medium and a dense, optically-thick 
component in the ``birth-clouds'' of actively star-forming regions, has been proposed to explain the excess 
attenuation suffered by the nebular emission \citep{calzetti94,charlot00}. Nebular emission is expected 
to arise from ionized gas located close to young, hot, ionizing stars within the dense birth-clouds (i.e.
is spatially associated with the optically-thick dust component), while stellar emission arises from stars 
both in such regions and throughout the disk; this is expected to be the reason for the higher extinction 
affecting the nebular emission.

While it is clear that nebular emission is significantly more extincted than the stellar continuum in local 
star-forming galaxies, the situation is much more unclear at high redshifts.  For example, similar extinction 
factors have been obtained for nebular line and stellar continuum emission in UV-selected galaxies at 
$z \approx 2$ \citep[e.g.][]{erb06,reddy10} and in far-IR-selected main-sequence galaxies at $z \approx 0.79-1.5$ 
\citep{puglisi16}, while higher extinction of the nebular 
emission (similar to or even larger than that in the local Universe) has been obtained in samples of optical- 
and near-IR-selected galaxies at $z \approx 1-2$ \citep[e.g.][]{forsterschreiber09,wuyts13,kashino13,price14}.
\citet{pannella15} also found evidence for redshift evolution in the differential extinction of nebular and 
stellar emission in their sample of sBzK galaxies, with similar extinction factors at high redshifts (out 
to $z \approx 3$) and higher extinction of the nebular emission (albeit by a lower factor, $\approx 1.3$, than 
that seen in the local Universe) at $z \approx 1$. Finally, given that it is possible that the results may 
depend on galaxy type, we note that \citet{price14} obtained a higher attenuation (by a factor of $\approx 1.8$) 
for the nebular emission in a sample of 163 main-sequence star-forming galaxies at $z \approx 1.36-1.5$.

In the previous section,  we estimated the dust extinction factors in the DEEP2 galaxies for three UV/optical SFR 
indicators, the nebular O{\sc ii}$\lambda$3727\AA\ line, and the stellar NUV ($\approx 230$~nm) and U-band 
($\approx 360$~nm) continua. Here, we compare the extinction of nebular emission with that of stellar emission 
in the DEEP2 galaxies; for this, we use the SFR estimate from the U-band continuum, as its wavelength 
($\approx 3600$~\AA) is very similar to that of the O{\sc ii}$\lambda$3727\AA\ line. The dust extinction factor 
$\epsilon(\lambda)$ is the ratio of the intrinsic luminosity $\rm F_{int}(\lambda)$ to the observed luminosity 
$\rm F_{obs}(\lambda)$, and is hence related to the color excess [$\rm E(B-V)$] and the reddening curve [$\rm k(\lambda)$]
by \citep[e.g.][]{calzetti00}
\begin{equation}
\rm \epsilon(\lambda) = \frac{F_{int}(\lambda)}{F_{obs}(\lambda)} \equiv 10^{\rm 0.4.E(B-V).k(\lambda)} \;.
\end{equation}
The ratio of the logarithms of the dust extinction factors for the O{\sc ii}$\lambda$3727\AA\ line 
emission and the U-band continuum is then 
\begin{equation}
\rm E_{OII/U} = \frac{log(\epsilon_{OII})}{log(\epsilon_U)} \equiv \frac{E_{gas}(B-V).k(373\;nm)}{E_{star}(B-V).k(360\;nm)} 
\approx \frac{E_{gas}(B-V)}{E_{star}(B-V)} \;,
\end{equation}
where we have assumed $\rm k(\lambda = 373\;nm) \approx k(\lambda = 360\;nm)$. We obtain $\rm E_{OII/U} \approx 
E_{gas}(B-V)/E_{star}(B-V) = (0.76 \pm 0.04)$ for the DEEP2 galaxies with $\rm M_B \leq -21$, over 
$0.7 <z  < 1.4$. The fact that $\rm E_{OII/U} < 1$ in our sample indicates that the nebular emission in 
the DEEP2 galaxies suffers {\it less} extinction than the stellar continuum at a similar wavelength. This is very 
different from the situation in nearby galaxies (where the ratio is $\approx 1.7$) and is also qualitatively 
different from the results obtained in earlier studies of high-$z$ galaxies, which found either 
higher extinction of the nebular emission \citep[e.g.][]{price14} or comparable extinctions for the nebular and stellar 
emission \citep[e.g.][]{forsterschreiber09,puglisi16}. 

We emphasize that our earlier caveat that the measured rest-frame U-band continuum luminosity 
might contain contributions from old stars does not affect the above conclusion. Contamination of the U-band 
continuum by emission from old stars would imply that our U-band SFR estimate is an upper limit, i.e.  that 
our estimate of the dust extinction factor for the U-band stellar continuum is a {\it lower} limit. This 
then implies that our estimate of $\rm E_{OII/U}$ is an upper limit to the true value. In other words, 
the possibility of contributions from an old stellar population to the U-band luminosity can only reduce 
the inferred value of $\rm E_{OII/U}$, and hence does not affect our conclusion that the nebular 
emission in the DEEP2 galaxies of our sample suffers less extinction than the stellar emission.

The three panels of Fig.~\ref{fig:o2u} plot the ratio $\rm E_{OII/U} \approx E_{gas}(B-V)/E_{star}(B-V)$
as a function of colour, stellar mass, and redshift. The right and middle panels of the figure show that higher 
values of $\rm E_{OII/U}$ are obtained in redder and more massive galaxies: nebular emission in redder and more 
massive galaxies thus appears to suffer larger extinction (relative to the stellar continuum) than similar emission in 
bluer and less massive galaxies \citep[see also][]{puglisi16}. Further, the O{\sc ii}$\lambda$3727\AA\ emission 
in the reddest and most massive galaxies shows significantly larger extinction than the U-band stellar 
continuum. Both of these are consistent with a scenario in which larger amounts of the second dust component 
\citep[assuming a two-component dust model similar to that of][]{charlot00} are present in the actively 
star-forming regions in dusty, massive galaxies, and that this second dust component is less prevalent or 
absent in blue galaxies.

Fig.~\ref{fig:o2u}[C] shows that the ratio $\rm E_{OII/U}$ increases with decreasing redshift in the DEEP2 
galaxies, over $1.4 \gtrsim z \gtrsim 0.7$. The dashed straight line shows the best linear fit to the data;
we obtain ${\rm E_{OII/U}} = (1.51\pm0.24) - (0.61\pm0.30)\times z$. Using this relation to extrapolate to 
lower redshifts, we obtain $\rm E_{OII/U}  \approx E_{gas}(B-V)/E_{star}(B-V) = (1.51 \pm 0.24)$ at 
$z = 0$. Interestingly, this is consistent with the estimates of the excess extinction (a factor of $\approx 1.7$) 
suffered by nebular emission in local star-forming galaxies \citep[e.g.][]{calzetti94,calzetti00,cidfernandes05}. 
In the context of the two-component dust model, this suggests that the optically-thick dust component steadily 
builds up in actively star-forming regions of main-sequence galaxies, causing a steady increase in the excess 
attentuation suffered by the nebular emission relative to the stellar continuum.

To address the possibility that the U-band continuum luminosity might not be a good tracer of the SFR, 
we have also carried out the analysis via a different approach, using the dust extinction factor estimated 
from the NUV 230~nm continuum luminosity. To estimate the dust extinction factor for the stellar continuum
at 373~nm (i.e. at the wavelength of the O{\sc ii}$\lambda$3727 line), we need to know the extinction curve of 
the DEEP2 galaxies. Since this is not known, we use a range of local extinction curves [$\rm k(\lambda)$], for 
the Milky Way \citep{cardelli89}, Large and Small Magellanic Clouds \citep[LMC and SMC, respectively;][]{gordon03} 
and local starburst galaxies \citep{calzetti94}. These yield $\rm E_{OII/373\;nm} \approx E_{gas}(B-V)/E_{star}(B-V) 
= (0.63 - 0.93)$, with the lowest and highest values for the extinction curves of the local starbursts and the SMC, 
respectively. Note that our result based on using the U-band luminosity as an SFR indicator lies in the 
middle of these estimates. We also continue to find that the ratio $\rm E_{OII/373\;nm}$ increases with decreasing 
redshift, over $1.4 \gtrsim z \gtrsim 0.7$. Extrapolating to lower redshifts for each assumed reddening curve, 
we obtain $\rm E_{OII/373\;nm} \approx 1.3 - 1.9$ at $z = 0$ (again with the lowest and highest values for 
the extinction curves of local starbursts and the SMC, respectively), broadly consistent with estimates of the 
excess extinction suffered by nebular emission in local galaxies. We thus find no evidence that our conclusions
might be affected by our use of the U-band continuum luminosity as an SFR indicator.

\subsection{The atomic gas depletion time scale}
\label{subsec:h1dep}

The timescale on which neutral gas is depleted by star formation, and its dependence on redshift, stellar 
mass, etc., is a subject of much interest in studies of galaxy evolution 
\citep[e.g.][]{schiminovich10,saintonge11b,saintonge13,tacconi13,genzel15,schinnerer16}. If star formation 
activity is to be maintained in a galaxy beyond its gas depletion time scale, the gas content must be 
replenished, probably by accretion of gas from the inter-galactic medium. Most studies of the gas depletion 
timescale in high-$z$, star-forming galaxies have focussed on the molecular gas, due to the difficulty in 
estimating atomic gas masses in galaxies at $z \gtrsim 0.25$, where it is very difficult to detect the weak 
\hii\ emission line. Molecular emission studies of star-forming galaxies have obtained typical gas depletion 
timescales of $\approx 0.5-1$~Gyr \citep[e.g.][]{saintonge11b,tacconi13,genzel15,schinnerer16} for 
main-sequence galaxies at $z \approx 0 - 4$. At high redshifts, the molecular gas depletion timescale shows 
only a weak dependence on redshift and stellar mass, but a strong dependence on the sSFR \citep{genzel15}.
However, at low redshifts, this timescale has been found to depend on stellar mass: the molecular gas 
depletion timescale is larger by a factor of $\approx 6$ in the highest-mass galaxies \citep{saintonge11b}. 
We note that a recent study of molecular gas in absorption-selected galaxies at $z \approx 0.7$ has found 
evidence for significantly longer molecular gas depletion timescales, $\gtrsim 10$~Gyr \citep{kanekar18}, 
very different from those seen in emission-selected star-forming galaxies. 

In the local Universe, \citet{saintonge11b} used the {\it COLD~GASS} galaxy sample, comprising of galaxies 
at $0.025 < z < 0.05$ with stellar mass $\rm M_\star \geq 10^{10} \; M_\odot$, to study the dependence of 
the atomic gas depletion time on galaxy properties. They found that the average atomic gas depletion 
timescale is $\approx 3$~Gyr (albeit with considerable scatter) for the {\it COLD~GASS} galaxies, with 
little dependence on parameters like the stellar mass, the sSFR, etc. \citep[see also][]{schiminovich10}. 
The molecular and atomic gas depletion timescales are similar in the highest-mass {\it COLD~GASS} 
galaxies, with red colours and high surface densities, while the molecular gas depletion timescale is 
nearly an order of magnitude lower than the atomic gas depletion timescale in low-mass galaxies, with 
high sSFR's and low stellar surface density \citep{saintonge11b}. Similarly, extending to lower stellar 
masses, $\rm M_\star \geq 10^9 \; M_\odot$, the galaxies of the {\it xGASS} ``representative sample'' of 
\citet{catinella18} have a median atomic gas depletion time of $\approx 5$~Gyr, for objects with detections 
of \hii\ emission.

In the case of the DEEP2 galaxies, \citet{kanekar16} stacked the \hii\ emission from 868 galaxies
at $z \approx 1.3$ to obtain an upper limit of $\rm M_{HI}(3\sigma) \leq 2.1 \times 10^{10} \; M_\odot$ 
on their average \hi\ mass. We have stacked the rest-frame 1.4~GHz radio continuum emission from the same 
868 galaxies, to obtain an SFR of $\rm (24.2 \pm 3.7) \; M_\odot$~yr$^{-1}$. Combining this with the 
upper limit of \citet{kanekar16} on the \hi\ mass then implies a $3\sigma$ upper limit of $\approx 0.87$~Gyr 
on the atomic gas depletion time, in main-sequence star-forming galaxies at $z \approx 1.3$. 
Interestingly enough, the atomic gas depletion timescale is comparable to the molecular gas depletion 
timescale ($\approx 0.7$~Gyr) in similar star-forming galaxies at a similar redshift \citep{tacconi13,genzel15}. 
The DEEP2 galaxies of the sub-sample of \citet{kanekar16} have stellar masses $\rm \gtrsim 10^{9} \; M_\odot$, 
similar to those of the {\it xGASS} sample, but have far shorter atomic gas depletion times. The short 
inferred gas depletion time of the DEEP2 sub-sample emphasizes the need for replenishment of the 
atomic gas content of high-$z$ main-sequence galaxies. Atomic hydrogen thus appears to be a transient 
phase in star-forming galaxies at high redshifts, with the conversion from the atomic to the molecular phase 
taking place on a timescale comparable to the timescale for the conversion of the molecular gas to stars.
This is consistent with the speculation of \citet{saintonge11b}, that, at high redshifts, the bottleneck 
for star formation does not lie in the conversion of atomic gas to molecular gas.

\citet{krumholz08,krumholz09a} carried out a theoretical modelling of the formation of molecular 
hydrogen from the atomic phase in galactic disks, considering the atomic to molecular transition 
to take place in a thin spherical layer separating a purely molecular region from a region with
negligible molecular fraction \citep[see also][]{krumholz13}, and finding reasonable agreement 
with observational data on local samples of galaxies. We use this model to examine conditions
in the DEEP2 galaxies that might yield the observed atomic and molecular gas depletion timescales 
of $< 0.87$~Gyr and $\approx 0.7$~Gyr, respectively. In chemical equilibrium, the atomic and molecular 
gas depletion timescales are related by 
\begin{equation}
\rm \Delta t_{HI} = \Delta t_{H_2} / R_{H_2} \;,
\end{equation}
where $\rm \Delta t_{HI}$ and $\Delta t_{H_2}$ are the atomic and molecular gas depletion timescales,
and $\rm R_{H_2}$ is the atomic-to-molecular mass ratio, with 
\begin{equation}
\rm R_{H_2} = f_{H_2}/[1 - f_{H_2}] \;,
\end{equation}
and $\rm f_{H_2}$ is the H$_2$ fraction.

Equations (10--12) of \citet{krumholz13} relate $\rm f_{H_2}$ to the total gas surface density $\Sigma_0$,
the clumping factor $\rm f_c$, and the metallicity. 
\citet{krumholz13} note that the gas clumping factor depends on the scale over which the surface 
density is averaged, with $\rm f_c \approx 1$ on scales of $\approx 100$~pc, and $\rm f_c \approx 5$ 
on scales of $\approx 1$~kpc. We will assume $\rm f_c \approx 5$ averaged over the DEEP2 galaxies.
Finally, we assume that the DEEP2 galaxies have, on average, solar metallicity.

For the DEEP2 galaxies, the gas depletion timescales are $\rm \Delta t_{HI} < 0.87$~Gyr, 
and $\rm \Delta t_{H_2} \approx 0.7$~Gyr \citep{tacconi13}, implying $\rm R_{H_2} \geq 0.8$.
Using this and $\rm f_c \approx 5$ in Equations~(10--12) of \citet{krumholz13} then yields an 
average total gas surface density of $\rm \Sigma_0 \gtrsim 6 \; M_\odot$~pc$^{-2}$. Overall, 
within the model of \citet{krumholz13}, the average gas surface density in the DEEP2 galaxies 
must be relatively high to account for the similar atomic and molecular gas depletion timescales.

\section{Summary}
\label{sec:disc}

We have carried out deep GMRT 610~MHz imaging of four sub-fields of the DEEP2 Galaxy Survey, 
achieving RMS noise values of $\approx 14-39\;\mu$Jy/Beam in the different fields. We have detected 
the stacked rest-frame 1.4~GHz radio continuum emission from a near-complete (M$_B \leq -21$) sample 
of 3698 blue star-forming galaxies at $0.7 \lesssim z \lesssim 1.45$, and use the stacked images to 
study the redshift evolution of the SFR, the sSFR, and the main sequence of star-forming galaxies. This 
is the first study of high-$z$ star-forming galaxies where such radio stacking has been carried 
out at frequencies close to the rest-frame 1.4~GHz frequency, so that assumptions about the slope of 
synchrotron emission do not affect our results. We obtain a median total SFR of 
$\rm (24.4 \pm 1.4) \; M_\odot$~yr$^{-1}$ for the 3698 galaxies of the sample. We also find that the 
stacked continuum 
emission is unresolved, with a transverse size $< 8$~kpc at the median sample redshift of 
$z_{\rm med} \approx 1.1$. We find that both the median SFR and the median sSFR decrease with 
decreasing redshift, with SFR~$\propto (1+z)^{1.98 \pm 0.50}$ and sSFR~$\propto (1+z)^{3.94 \pm 0.57}$ 
over $0.7\lesssim z \lesssim 1.45$, consistent with earlier studies based on other SFR indicators.
We clearly detect the main-sequence relation between SFR and stellar mass, with with SFR~
$\rm = (13.4 \pm 1.8) \times \left[M_\star/(10^{10} \;  M_\odot)\right]^{(0.73 \pm 0.09)}$~M$_\odot$~yr$^{-1}$. 
We also find weak evidence that the normalization 
of the main sequence increases by a factor of $\approx 1.6$ from $z \approx 0.85$ to $\approx 1.2$; 
however, we find no evidence for changes in the main sequence slope over this redshift range. 
We compare the median SFRs estimated from other indicators, the [O{\sc ii}]$\lambda$3727 line 
luminosity, the rest-frame NUV continuum luminosity, and the rest-frame U-band continuum luminosity, 
with the median total SFR inferred from our stacking analysis
to infer the dust extinction correction factor for each tracer. We obtain dust extinction correction 
factors of $\approx 4.7$, $\approx 2.0$, $\approx 2.5$, and $\approx 2.7$ for the above 
four tracers, respectively, with the largest correction factor for the NUV luminosity and the 
smallest for the [O{\sc ii}]$\lambda$3727 line luminosity. This indicates that significant 
dust extinction is present in the DEEP2 galaxies. We find that the dust extinction correction factors 
do not appear to vary with redshift, over $0.7 \lesssim z \lesssim 1.45$, but increase with 
increasing color and stellar mass, for all SFR tracers. Nebular emission appears to suffer less 
extinction than the stellar continuum in the DEEP2 galaxies, contrary to the situation in 
galaxies in the local Universe and a few studies of galaxies at high redshifts. However, the relative 
extinction increases with both increasing stellar mass and colour, with redder, more massive galaxies 
showing higher extinction in the nebular emission than in the stellar continuum. This suggests that 
the actively star-forming regions in red, massive galaxies at $z \approx 1$ already contain significant 
amounts of a second dust component, different from that in the extended galactic disk. We also find 
that the ratio of nebular extinction to stellar extinction in the DEEP2 galaxies increases with decreasing 
redshift; extrapolating this relation to $z = 0$, we find that this ratio is consistent with estimates of 
the excess extinction suffered by nebular emission relative to the stellar continuum in nearby 
star-forming galaxies. Finally, we combine our median radio SFR estimates with the upper limit 
on the average \hi\ mass of a sub-sample of the DEEP2 galaxies at $z \approx 1.3$ to obtain an 
upper limit of $0.87$~Gyr on the atomic gas depletion time for star-forming galaxies at this 
redshift. This is the first constraint on the atomic gas depletion 
time in star-forming galaxies at $z \gtrsim 1$; the low value of the gas depletion time suggests 
that \hi\ is likely to be a transient phase in star-forming galaxies, with efficient conversion 
of atomic gas to molecular gas, on a timescale similar to that of the conversion of the molecular 
gas to stars.

\acknowledgments
We thank the GMRT staff who have made these observations possible. The GMRT is run by the 
National Centre for Radio Astrophysics of the Tata Institute of Fundamental Research. 
NK acknowledges support from the Department of Science and Technology via a Swarnajayanti 
Fellowship (DST/SJF/PSA-01/2012-13). AB thanks Jayaram Chengalur and Prasun Dutta for 
useful discussions. We thank Mark Krumholz for discussions and much advice about the 
application of his model to our results, and an anonymous referee for his/her detailed 
comments and suggestions on an earlier version of the manuscript.

\bibliographystyle{apj}

\end{document}